\documentclass[sigconf,screen]{acmart}

\usepackage{algorithm}
\usepackage[noend]{algpseudocode}

\usepackage{enumitem}
\usepackage{multirow}
\usepackage{makecell}
\usepackage{xcolor}
\usepackage{microtype}

\newif\ifshowblue
\showbluefalse   % 【关蓝色】取消注释=蓝色变黑

\newif\ifshowred
\showredfalse    % 【关红色】取消注释=红色变黑

\newcommand{\verified}[1]{\ifshowblue\textcolor{blue}{#1}\else#1\fi}
\newcommand{\important}[1]{\ifshowred\textcolor{red}{#1}\else#1\fi}
\let\oldtexttt\texttt
\renewcommand{\texttt}[1]{{\small\oldtexttt{#1}}}

\usepackage{ifthen}

\newboolean{showchangesformajor}
\setboolean{showchangesformajor}{false} % <--- Toggle this to control global visibility

\usepackage{xcolor}
\newenvironment{revision}{%
    \ifthenelse{\boolean{showchangesformajor}}%
        {\color{blue}}% Set text color to blue if switch is true
        {}%            Do nothing (keep default black) if switch is false
}{%
    \ignorespacesafterend
}

\newcommand{\rev}[1]{\ifthenelse{\boolean{showchangesformajor}}{\textcolor{blue}{#1}}{#1}}

\AtBeginDocument{%
  }

\copyrightyear{2026}
\acmYear{2026}
\setcopyright{cc}
\setcctype{by}
\acmConference[ASE '26]{Proceedings of the 41st IEEE/ACM International Conference on Automated Software Engineering}{October 12--16, 2026}{Munich, Germany}
\acmBooktitle{Proceedings of the 41st IEEE/ACM International Conference on Automated Software Engineering (ASE '26), October 12--16, 2026, Munich, Germany}
\acmDOI{10.1145/3832783.3837474}
\acmISBN{979-8-4007-2882-2/2026/10}

\begin{document}
\hypersetup{colorlinks=true,linkcolor=blue,urlcolor=blue,citecolor=black}

\title{We Must Have Missed This Comment:
Detecting and Repairing Stale Function References in Linux Kernel Comments}
% \title{\textit{``We Must Have Missed This Comment''}:
% Detecting and Repairing Stale Function References in Linux Kernel Comments}
% \title{Outdated Comments Can Even Affect Code: \\
% Identify Outdated Comments with Patching in Linux kernel}
% \title{Outdated Comments can even affect Code: \\
% Identify Outdated comments with Patching in Linux kernel}

\author{Kexin Sun}
\orcid{0009-0004-1344-5444}
\email{kexinsun@smail.nju.edu.cn}
\affiliation{%
  \institution{State Key Lab for Novel Software Technology, Nanjing University}
  \city{Nanjing}
  \country{China}}

\author{Yunbo Lyu}
\orcid{0009-0004-2522-7348}
\authornote{Hongyu Kuang and Yunbo Lyu are the corresponding authors.}
\email{yunbolyu@smu.edu.sg}
\affiliation{%
  \institution{Singapore Management University}
  % \city{Singapore}
  \country{Singapore}}

\author{Xutong Ma}
\orcid{0009-0009-1510-1668}
\email{xutong.ma@inria.fr}
\affiliation{%
  \institution{Inria}
  \city{Paris}
  \country{France}}

\author{Hongyu Kuang}
\authornotemark[1] 
\orcid{0009-0003-8702-2826}
\email{khy@nju.edu.cn}
\affiliation{%
  \institution{State Key Lab for Novel Software Technology, Nanjing University}
  \city{Nanjing}
  \country{China}}

\author{Ratnadira Widyasari}
\orcid{0000-0001-8190-5458}
\email{ratnadiraw@smu.edu.sg}
\affiliation{%
  \institution{Singapore Management University}
  % \city{Singapore}
  \country{Singapore}}

\author{He Zhang}
\orcid{0000-0002-9159-5331}
\email{hezhang@nju.edu.cn}
\affiliation{%
  \institution{State Key Lab for Novel Software Technology, Nanjing University}
  \city{Nanjing}
  \country{China}}

\author{Xiaoxing Ma}
\orcid{0000-0001-7970-1384}
\email{xxm@nju.edu.cn}
\affiliation{%
  \institution{State Key Lab for Novel Software Technology, Nanjing University}
  \city{Nanjing}
  \country{China}}

\author{Julia Lawall}
\orcid{0000-0002-1684-1264}
\email{julia.lawall@inria.fr}
\affiliation{%
  \institution{Inria}
  \city{Paris}
  \country{France}}

\author{David Lo}
\orcid{0000-0002-4367-7201}
\email{davidlo@smu.edu.sg}
\affiliation{%
  \institution{Singapore Management University}
  % \city{Singapore}
  \country{Singapore}}

\renewcommand{\shortauthors}{Sun et al.}

\begin{abstract}
As the Linux kernel evolves, code comments may become outdated, as the functions they reference can be refactored or removed independently without corresponding updates to the comments. Such \emph{stale function references} can mislead maintainers and thus hinder code comprehension.
Prior work on detecting code-comment inconsistency mainly focused on addressing semantic misalignment between Javadoc comments and their directly annotated functions, making them inapplicable to this type of \emph{externally induced} staleness in the Linux kernel.
Therefore, we propose \textsc{ReCite}, a three-stage approach to identify and repair such stale references: (1)~detecting unresolved \emph{function-form symbols}---symbols in comments that appear to reference functions but for which no matching function can be found in the current codebase, (2)~tracing the evolution history of each unresolved symbol through the Git history, and (3)~generating LLM-based repair suggestions grounded in the evolution history and current code context.
On Linux kernel v6.18-rc1, \textsc{ReCite} detects \verified{869} stale references with generated repair suggestions.
A manual evaluation on 200 sampled repairs shows that 178 (89.0\%) provide useful repair guidance, with 85 (42.5\%) directly applicable.
Of our 75 submitted patches, 50 have been accepted.
We also empirically study all unresolved function-form symbols.

% We have submitted 75 patches to the Linux kernel and \important{50} have been accepted.
% Additionally, we empirically study all unresolved function-form symbols located in Linux Kernel comments.
\end{abstract}




\maketitle

\section{Introduction \& Motivation}
% Project-level
% We have xx pathces been accepted by Linux kernel.
% Ctag has been used in~\cite{lyu2024evaluating}

In modern iterative development, code undergoes frequent refactoring as part of routine maintenance~\cite{silva2016we,fowler2018refactoring,ivers2022industry}.
The Linux kernel, which is one of the world's most actively maintained codebases, typically ships every two to three months with each cycle absorbing about 13,000 changesets~\cite{linux-kernel-process}.
Such changes, however, do not always extend to the comments: 
% CITE: lyu2024evaluating
existing automated refactoring approaches are primarily proposed for syntactic code elements, instead of unstructured comment texts, making comments easily outdated during code evolution~\cite{ratol2017detecting}.
The lack of update to relevant comments has also been substantiated by a large-scale empirical study~\cite{wen2019large}.
Within these outdated comments, a particularly harmful case is the \emph{stale function reference}, where the comment of a given function typically mentions another function that has been refactored or removed and thus no longer exists in the current codebase.
Such references matter because they capture repository-level knowledge from the original developers that links code entities across the codebase.
Once stale, they reduce comment usefulness and increase the maintenance burden as the codebase evolves~\cite{tan2007icomment,ibrahim2012relationship}.

% Prior studies~\cite{huang2025your,rong2025code} have recognized the code comment inconsistency (CCI) issue.
% Unfortunately, the majority of existing work concentrates on code elements and their corresponded Javadoc comments in Java~\cite{panthaplackel2021deep}.
% This situation makes it hard to apply existing CCI approaches for detecting stale function references in Linux Kernel comments because: (1) they did not support C-language codebases such as the Linux kernel; (2) these approaches mainly focus on the structural pairing between each Javadoc comment and its annotated function.
% The latter situation is inherently limited to local semantic inconsistencies and cannot be detected when a referenced function evolves independently outside the commenting code.
% An illustrating sample for further demonstration is depicted in Figure~\ref{fig:difference}.
% Particularly, we focus on \textit{externally induced} stale function references, where a comment becomes outdated because the function it mentions has been refactored or removed independently elsewhere.

\begin{figure}[t]
    \centering
    \setlength{\abovecaptionskip}{6pt}
    \includegraphics[width=0.38\textwidth]{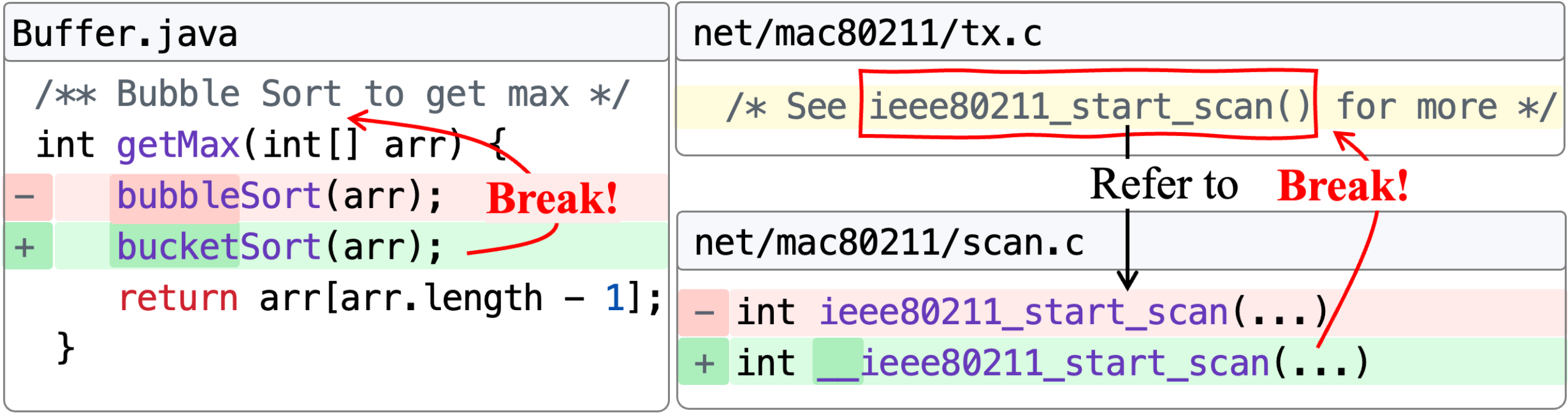}
    \caption{
    % Prior CCI work vs.\ our focus. (a)~Prior work detects broad local semantic misalignment. (b)~We target stale function references caused by external evolution.
    Prior CCI work vs. our focus. Prior work detects broad local semantic
  misalignment (left). We target stale function references caused by external
  evolution (right).}
    \label{fig:difference}
  \end{figure}

\begin{revision}
Prior studies~\cite{huang2025your,rong2025code} have recognized the code comment inconsistency (CCI) issue, but they generally focus on \textit{local} inconsistencies between a function and its directly associated comment, typically Javadoc-style function-comment pairs in Java~\cite{panthaplackel2021deep} (Figure~\ref{fig:difference}, left).
In contrast, we target a different type of inconsistency that we observe in Linux kernel comments: \textit{externally induced} stale function references, where a comment becomes outdated because the function it mentions has been refactored or removed independently elsewhere (Figure~\ref{fig:difference}, right).
Such inconsistencies are overlooked by approaches that focus only on local inconsistencies.
\end{revision}

\begin{figure}[t]
    \centering
    \setlength{\abovecaptionskip}{6pt}
    \includegraphics[width=0.42\textwidth]{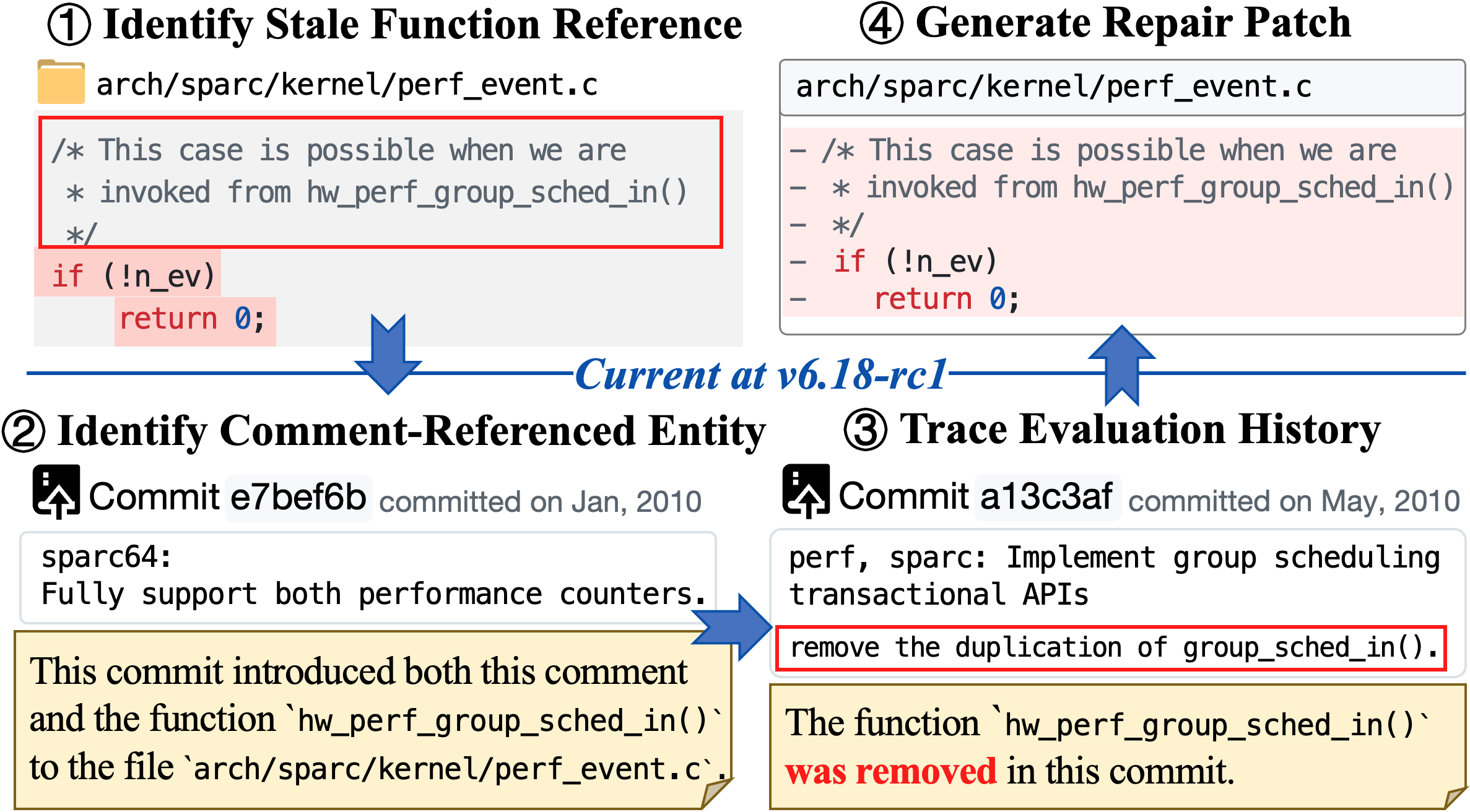}
    \caption{A motivating example of a stale function reference to \texttt{hw\_perf\_group\_sched\_in()} and its repair.}
    \label{fig:motivation_example}
\end{figure}

\begin{revision}
These stale references can affect more than comment readability.
Figure~\ref{fig:motivation_example} shows a case where a comment describing a conditional exit refers to  \texttt{hw\_perf\_group\_sched\_in()}, a function that no longer exists in the current codebase.
To repair this unresolved reference, we traced back to commit \href{https://github.com/torvalds/linux/commit/e7bef6b04ca2e8e4cf667c43d7e2ab3034a869d5}{\texttt{e7bef6b}} (January 2010), where the comment was originally written, and found that \texttt{hw\_perf\_group\_sched\_in()} was defined at line 992 of the same file.
We then tracked the function's subsequent evolution and found that it was completely removed in commit \href{https://github.com/torvalds/linux/commit/a13c3afd9b62b6dace80654964cc4ca7d2db8092}{\texttt{a13c3af}} (April 2010).
This removal not only rendered the comment stale, but also left the conditional branch ``\texttt{if (!n\_ev) return 0;}'' as unreachable dead code.
Such dead code unnecessarily inflates the codebase that maintainers and security auditors must reason about---a well-recognized concern in software debloating and kernel attack-surface reduction~\cite{kurmus2013attack, quach2018debloating}.
We have submitted a patch that removes both the outdated comment and the dead code.
This case illustrates that stale function references are not merely a documentation problem---they can signal latent logical defects that may otherwise go unnoticed.
\end{revision}

\begin{revision}
Linux kernel developers have also noticed such stale references in practice.
% For example, in commit \href{https://github.com/torvalds/linux/commit/b7dd80f8f92848fa26518119f2c378dad8b7c0da}{\texttt{b7dd80f}} (November 2025), a developer manually updated two comments to remove stale references to \texttt{scan\_swap\_map\_slots()}— a function that had been removed earlier in commit \href{https://github.com/torvalds/linux/commit/0ff67f990bd45726e0d9e91111d998e7a3595b32}{\texttt{0ff67f9}} (March 2025).
For example, commit \href{https://github.com/torvalds/linux/commit/b7dd80f8f92848fa26518119f2c378dad8b7c0da}{\texttt{b7dd80f}} removed two references to the deleted \texttt{scan\_swap\_map\_slots()} from comments.
\end{revision}
Nevertheless, such fixes remain isolated and manual, and the kernel still lacks a systematic approach to (i) scan the codebase for stale function references, (ii) trace when and why a reference became \emph{unresolved} (i.e., no matching function can be found for the symbol in the codebase), and (iii) generate history-grounded repairs at scale.\looseness=-1

To bridge this gap, we propose \textsc{ReCite}, a three-stage approach for automatically identifying and repairing stale function references in Linux kernel comments:
(1) Given a kernel snapshot, \textsc{ReCite} first extracts \emph{function-form symbols} (i.e., symbols written as \texttt{``foo()''} that appear to reference a function or a function-like macro) from all comment blocks.
It then checks whether they are resolvable (i.e. the matching function can be found) in the current codebase, using \texttt{grep} combined with Coccinelle~\cite{padioleau2008documenting}, a semantic patching framework for C code.
% (2) For each unresolved symbol, \textsc{ReCite} traces back to the commit where the reference was introduced and locates the function definition at that time via large language model (LLM)-based entity selection.
\begin{revision}
(2) For each unresolved symbol, \textsc{ReCite} traces back to the commit where the reference was introduced and identifies the function actually intended by the comment via large language model (LLM)-based entity selection, since static analysis alone cannot disambiguate among same-name candidates.
\end{revision}
It then tracks the function's evolution forward---through renames, splits, and removals---to construct an evolution tree.
(3) Finally, \textsc{ReCite} prompts an LLM with the evolution tree and relevant context to generate a history-grounded repair suggestion.

We evaluate \textsc{ReCite} on Linux kernel v6.18-rc1, where it detects 869 stale references and automatically generates corresponding repair suggestions.
\begin{revision}
A manual evaluation on 200 sampled generated repairs shows that 178 (89.0\%) provide useful repair guidance, including \textbf{85 (42.5\%) directly applicable} repairs and 93 (46.5\%) partially acceptable ones that need minor manual refinement before direct application.
\end{revision}
\begin{revision}
We therefore position \textsc{ReCite} as a decision-support tool that provides helpful repair candidates for maintainers, rather than a fully automatic repair system, at present.
\end{revision}
We have submitted 75 patches to the Linux kernel; \textbf{\important{50} have been accepted}.
Specifically, a maintainer who received our patch remarked: ``\textit{We must have missed this comment ... a few years ago.}''
We further did an empirical study on all detected unresolved function-form symbols (5,442 in total) from 8,258 mentions, i.e., occurrences in comments, and established a five-category taxonomy of their underlying causes.\looseness=-1

\section{Related Work}

% We organize related work into two areas relevant to our study: (a) code-comment inconsistency detection and (b) code--documentation inconsistency across software artifacts.

% \vspace{0.5em}
% \noindent
\textbf{Code Comment Inconsistency.}
% Early work on code-comment inconsistency relies on rule-based approaches.
% Ratol and Robillard~\cite{ratol2017detecting} define lexical rules to detect \emph{fragile comments} whose correctness depends on specific identifiers, reporting matches as Eclipse warning markers.
% Early work on code-comment inconsistency relies on rule-based approaches to detect fragile comments.
Early work relies on rule-based approaches to detect code-comment inconsistency~\cite{tan2007icomment,tan2011acomment,tan2012tcomment,ratol2017detecting},
while subsequent work shifts towards learning-based methods~\cite{liu2018automatic, rabbi2020detecting, xu2024code}
For example, Panthaplackel et al.~\cite{panthaplackel2021deep} use a GNN-based encoder with attention to jointly represent code edits and comments, predicting whether a change invalidates the associated comment at commit time.
Other learning-based approaches include CoCC~\cite{huang2025your}, which uses skip-gram embeddings and random-forest classifiers for binary judgments, 
and CARL-CCI~\cite{nguyen2026carl}, which applies label-aware contrastive learning on structured code diffs.
More recently, the task has expanded from detection to automated repair~\cite{panthaplackel2020learning, liu2020automating, lin2023cct5}, as demonstrated by Rong et al.~\cite{rong2025code}, who fine-tune CodeLLaMA (C4RLLaMA) to both detect and repair inconsistencies.
Compared to prior CCI work, we target a long-overlooked problem: a comment may reference a function in a distant file that evolves independently, leaving the comment stale.
Existing CCI approaches, developed mostly for Java documentation comments, target local function-summary inconsistencies and thus miss cross-file references (60.5\% of our cases; Section~\ref{sec:entity-selection-results}). 
Our C4RLLaMA evaluation further indicates poor transfer to this setting (Section~\ref{sec:rq4}).

% Existing CCI approaches, mostly studied on Java-style documentation comments, typically check local inconsistency between a function and its own summary, and thus cannot capture references that cross files (60.5\% of our cases; see Section~\ref{sec:entity-selection-results}).
% Our evaluation with C4RLLaMA~\cite{rong2025code} further suggests that such general CCI approaches do not transfer well to this setting (see Section~\ref{sec:rq4}).

% All of these studies target Java and rely on local co-change signals---linking a Javadoc comment to the method it directly documents.
% In contrast, our work addresses C-language comments in the Linux kernel, where a comment may reference a function defined in a distant file.
% Such \emph{externally induced} stale references cannot be captured by local co-change heuristics.
% To validate this gap, we evaluated C4RLLaMA on 38 manually confirmed cases (sampled from our dataset) where direct name substitution is the most appropriate repair for stale function references.
% Although C4RLLaMA detected 31 as inconsistent, none of its outputs suggested the correct replacement function name.
% This motivates \textsc{ReCite}, which addresses this gap through cross-file symbol resolution and evolution-history tracing to generate history-grounded repairs.]
% Given this mismatch in problem scope, we do not include prior CCI approaches as baselines.
%%%%

\begin{revision}
% Compared to prior CCI work, we target a long-overlooked problem: a comment may reference a function in a distant file that evolves independently, leaving the comment stale.
% Existing CCI approaches, mostly studied on Java-style documentation comments, typically check local inconsistency between a function and its own summary, and thus cannot capture references that cross files (60.5\% of our cases; see Section~\ref{sec:entity-selection-results}).
% Our evaluation with C4RLLaMA~\cite{rong2025code} further suggests that such general CCI approaches do not transfer well to this setting (see Section~\ref{sec:rq4}).
\end{revision}

\begin{figure*}[t]
    \centering
     \setlength{\abovecaptionskip}{6pt}
    \includegraphics[width=0.67\textwidth]{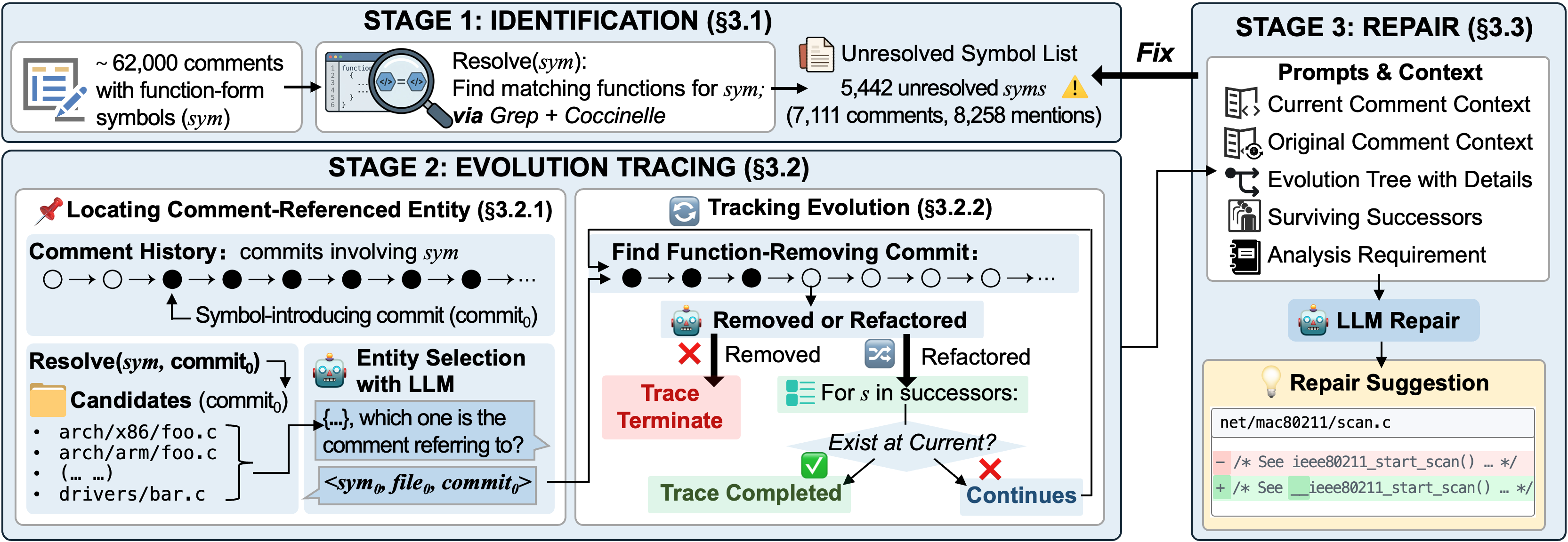}
    \caption{Overview of \textsc{ReCite} for identifying and repairing
  stale function references in Linux kernel comments.}
    \label{fig:pipeline}
\end{figure*}

% \vspace{0.5em}
% \noindent
\textbf{Code--Documentation Inconsistency Across Software Artifacts.}
Beyond source-code comments, inconsistencies between evolving code and its accompanying documentation are pervasive across software artifacts.
Prior work has studied outdated code names or samples in API documentation~\cite{zhong2013detecting,lee2019automatic,lee2025can},
stale code references in README files~\cite{tan2024detecting,tan2023wait} and outdated code snippets on Q\&A posts~\cite{ragkhitwetsagul2019toxic}.
With the growing capability of large language models, recent work has also begun applying LLMs to detect and repair such inconsistencies.
For example, METAMON~\cite{lee2025metamonfindinginconsistenciesprogram}, DocPrism~\cite{xu2025docprismlocalcategorizationexternal}, and Gao et al.~\cite{gao2026doesreadmefileneed} use LLM-based reasoning to detect broader forms of code--documentation inconsistency across code behavior, documentation text, and README updates.
% \begin{revision}
These efforts share our concern for code--documentation inconsistency, but they target different scenarios.
Tang et al.~\cite{tang2026detecting} target outdated screenshots, 
while Lee et al.~\cite{lee2019automatic,lee2025can} and Tan et al.~\cite{tan2024detecting,tan2023wait} target outdated API references in project documentation rather than in-code comments, with documentation-specific designs.
Beyond this artifact gap, most of these works (e.g. ~\cite{zhong2013detecting, lee2025metamonfindinginconsistenciesprogram}) reason over a single code snapshot or a single commit diff, and can thus only establish whether an inconsistency exists.
Our approach traces full evolution histories to determine not only \emph{whether} a reference is stale, but also \emph{why} it became stale and \emph{how} to repair it.
FreshDoc is closest in using revision history for API documentation, but its Java-AST-based change analysis~\cite{fluri2007change} cannot handle C constructs such as Linux kernel \#ifdef directives.
We therefore adopt RefDiff~\cite{silva2020refdiff} as a baseline for evaluating \textsc{ReCite}'s LLM-based evolution reasoning.

% Our approach instead traces the full evolution history, enabling it to determine not only \emph{whether} a reference is stale, but also \emph{why} it became stale and \emph{how} to repair it.
% FreshDoc is the closest prior work in its use of revision history to detect outdated references in API documentation;
% however, its core change-analysis algorithm~\cite{fluri2007change} is Java-AST-bound and blind to C-specific constructs in the Linux kernel (e.g., \#ifdef directives), making it hard to transfer to our setting.
% We therefore adopt a more advanced change-analysis baseline (RefDiff~\cite{silva2020refdiff}) in experiments to demonstrate the accuracy of \textsc{ReCite}'s LLM-based reasoning over evolution history.
% \end{revision}

\section{Proposed Approach}
Inspired by the manual repair process, we propose \textsc{ReCite}, a three-stage approach (Figure~\ref{fig:pipeline}) for automatically identifying and repairing stale function references in Linux kernel comments.
First, we scan the codebase to extract all function-form symbols from comments, and determine which of them are unresolved in the current codebase (Section~\ref{sec:identification}).
% We then analyze why these references are unresolved.
% Generally, we classify them into two categories: those that were resolvable (i.e., the function existed) when the comment was written, and those that were not.
% For unresolvable cases, we conduct a manual analysis to categorize their underlying reasons (detailed in Section~\ref{sec:results}).
% For the resolvable cases—those that became stale purely due to code evolution—we trace their complete evolutionary history in the codebase (Section~\ref{sec:history}).
We then trace the evolutionary history of each unresolved symbol to understand why it became invalid (Section~\ref{sec:history}).
Finally, we leverage an LLM to generate repair suggestions based on the function's evolutionary history (Section~\ref{sec:repair}).

% \item \textbf{Tree-sitter~\cite{tree-sitter}}:
% a widely used parser that builds concrete syntax trees from source code.
% We query ``\texttt{function\_definition}'', ``\texttt{function\_declarator}'', and ``\texttt{preproc\_function\_def}'' nodes to check whether the identifier is defined or declared.

% \item \textbf{Universal Ctags~\cite{universal-ctags}}:
% a syntax-based tool that generates a per-file symbol index.
% We invoke it with ``\texttt{--kinds-C=+f+d+p}'' to extract function definitions (``\texttt{f}''), macro definitions (``\texttt{d}''), and function declarations (``\texttt{p}''), then check whether the target identifier appears in the resulting index.

% \item \textbf{GNU Global~\cite{gnu-global}}:
% a project-wide navigation tool that builds a cross-reference database.
% We construct an index for each candidate file and query it for matching definitions (``\texttt{global -d}'') and symbol references (``\texttt{global -s}''), then verify each hit against syntactic patterns to confirm it corresponds to a function definition, declaration, or function-like macro.

\subsection{Identifying Unresolved Function-Form Symbols}
\label{sec:identification}
% We anchor our analysis on Linux kernel v6.18-rc1 (\texttt{HEAD}=\href{https://github.com/torvalds/linux/commit/3a8660878839faadb4f1a6dd72c3179c1df56787}{\texttt{3a86608}}).
% From around 62,000 source files (\texttt{.c} and \texttt{.h}), we extract \verified{2.27} million code comments, including both block comments (``\texttt{/* \ldots\ */}'') and line comments (``\texttt{// \ldots}'').
% From these, we identify \emph{function-form symbols}---tokens matching the regular expression
% ``\texttt{[A-Za-z\_][A-Za-z\\0-9\_]*()}''.
% This yields \verified{49,302} unique symbols across \verified{61,866} comments, totaling \verified{83,241} mentions.
We analyzed Linux kernel v6.18-rc1 (\texttt{HEAD}=\href{https://github.com/torvalds/linux/commit/3a8660878839faadb4f1a6dd72c3179c1df56787}{\texttt{3a86608}}).
From  around 62,000 source files (\texttt{.c} and \texttt{.h}), we extracted 2.27 million code comments, and identified \emph{function-form symbols} as tokens matching the regular expression
``\texttt{[A-Za-z\_][A-Za-z0-9\_]*()}''.
This yields \verified{49,302} unique symbols across \verified{61,866} comments, totaling \verified{83,241} mentions.
\begin{revision}
 Here, a \emph{symbol} refers to a unique function-form string, while a \emph{mention} refers to one occurrence of such a symbol in a comment, as one symbol can appear in multiple comments.
 % (or multiple times in one comment).
\end{revision}
We then design a two-step validation pipeline to determine whether a given symbol has at least one matching function-level code entity (i.e., can be ``\emph{resolved}'')  at a specific snapshot.\looseness=-1

\textbf{Step 1: Coarse-Grained Search.}
% For each symbol, we first obtain candidate source files that mention it, using \texttt{grep} (i.e., ``\texttt{grep -n -F -w <symbol> <commit> -- *.c *.h}'').
For each symbol, we use \texttt{grep} to obtain candidate source files that mention it.
We then filter out matches that occur exclusively within comment regions or string literals, ensuring the matched identifier appears in code proper.

\textbf{Step 2: Fine-Grained Static Analysis.}
For symbols with candidate files, we further apply static analysis to check whether a matching function definition, declaration, or function-like macro definition exists.
% We include function declarations as valid evidence of existence, because some functions are implemented in assembly (e.g., \texttt{secondary\_startup\_arm} in \texttt{arch/arm/kernel/head.S}) and only declared in C files.
We include function declarations as valid evidence of existence, because some functions are implemented in assembly and only declared in C header files.
Ignoring declarations would over-approximate the unresolved symbols.
We adopt \textbf{Coccinelle}~\cite{padioleau2008documenting} for this check.
Coccinelle is a code matching and patching engine based on the Semantic Patch Language
(SmPL), which was originally developed to support collateral evolution in the
Linux kernel.
We write an SmPL rule that matches function definitions of the form ``\texttt{name(...)\{...\}}'', ``\texttt{type name(...);}'' and ``\texttt{\#define name(...) ...}''.
% We write an SmPL rule that matches function definitions of the form ``\texttt{name(...)\{...\}}''.
% For function declarations and function-like macros, where the kernel's pervasive use of non-standard type qualifiers and compiler attributes can disrupt SmPL's semantic matching, we supplement with explicit syntactic pattern matching for ``\texttt{type name(...);}'' and ``\texttt{\#define name(...) ...}''.

We denote this two-step pipeline as \textsc{Resolve}(\textit{sym},\ \textit{commit}), which returns the set of files that contain the matching function-level code entity for \textit{sym} in version \textit{commit}.
When restricted to a single file, we write \textsc{Exists}(\textit{sym}, \textit{file}, \textit{commit}) to denote whether \textit{sym} can be resolved within a specific file.
If no matching entity is found in any candidate file, the symbol is considered unresolved at the queried version.
We treat the unresolved symbols in snapshot v6.18-rc1 as the starting point for the subsequent repair pipeline.

\subsection{Tracing Function Evolution History}
\label{sec:history}

To understand why a function-form symbol is unresolvable in the current snapshot, we trace its evolutionary history.
We model the historical state of a code entity as a tuple $\langle \textit{sym},\ \textit{file},\ \textit{commit} \rangle$, where
\textit{sym} is the function-form symbol,
\textit{file} is the source file path, and
\textit{commit} is the Git commit hash.
The goal of this stage is to first construct the initial state $\langle \textit{sym}_0,\ \textit{file}_0,\ \textit{commit}_0 \rangle$ by locating the code entity that the comment originally referred to (Section~\ref{sec:locating}), and then trace its evolution forward along the Git history (Section~\ref{sec:tracking}).

\subsubsection{Locating the Referenced Code Entity}
\label{sec:locating}

We first return to the commit where the symbol was originally introduced into the comment (i.e., the \emph{symbol-introducing commit}), and verify whether it was resolved at that point.
For the symbols that are already unresolvable at that time, we manually categorize the underlying reasons in Section~\ref{sec:taxonomy}.
For the resolvable symbols, we further identify which specific code entity the comment refers to.

\textbf{1) Identify the symbol-introducing commit}:
% We first retrieve the rename history of the file containing the comment via ``\texttt{git log --follow --name-status}, then use \texttt{git log -m -G <symbol>}'' on each historical file path to collect commits whose patches involve the symbol.
% We first retrieve the rename history of the commented file, and collect commits whose patches involve the referenced symbol under each historical file path.
Since files may be renamed over time, we first trace the rename history of the commented file.
We then collect commits whose patches involve the referenced symbol under each historical file path.
Starting from the earliest such commit, we check whether any comment-region line in the file at that version has a Jaccard similarity $\geq 0.7$ with the symbol-bearing line in the current comment.
We use Jaccard similarity rather than exact matching because comments are frequently extended or lightly edited. The 0.7 threshold tolerates minor wording changes while reliably identifying the original introduction.
The earliest commit satisfying this criterion is taken as the symbol-introducing commit $\textit{commit}_0$.

\textbf{2) Collecting Candidate Code Entities}:
We collect all candidate files containing a matching function entity at the symbol-introducing commit by invoking \textsc{Resolve}(\textit{sym}, $\textit{commit}_0$).

\textbf{3) Selecting the Comment-Referenced Code Entity with the LLM}:
The static analysis alone cannot determine which candidate code entity the comment actually refers to~\cite{lyu2025my}.
The Linux kernel supports multiple architectures, and the same function name may have several architecture-specific implementations simultaneously (e.g., under x86 and ARM).
Moreover, some function-form symbols do not refer to any function at all---for instance, ``\texttt{decrypt()}'' in the comment ``\textit{an optional alternate error code that is acceptable for \texttt{decrypt()} to return}'' refers generically to a decryption operation rather than a specific implementation.
Therefore, we leverage the LLM to identify which candidate the comment refers to.

% We supply the LLM with: the target symbol; the comment text at the symbol-introducing commit; the comment-related code context; and the set of candidate entities with their surrounding code.
% To obtain comment-related context, we parse the source file with Tree-sitter~\cite{tree-sitter} and extract function blocks based on \texttt{function\_defi\-nition} and \texttt{preproc\_function\_def} nodes (i.e., function-like macro definitions).
% Specifically:
% 1) If the comment lies inside one of these blocks, we use the entire block as context;
% 2) If the comment is near a block (within three lines), we use the nearest block as the context;
% 3) As a fallback, we expand bidirectionally up to 10 lines centered on the comment.
% Regarding the candidate entities, we represent each candidate by its identifier line together with the three preceding lines.
% Because the function signatures may span multiple lines, and qualifiers such as \texttt{static}, which affect a function's visibility (e.g., file-local scope), may appear on a line preceding the function name.
% These candidates are ordered by path edit distance to the comment's file,
% with closer files listed first.
We supply the LLM with the target symbol, the comment text at the symbol-introducing commit, the comment-related code context, and the set of candidate entities.
The code context is extracted via Tree-sitter~\cite{tree-sitter} from the enclosing or nearest function.
Candidates are ordered by path edit distance to the comment's file.
Each candidate is represented by its identifier line together with the three preceding lines, as function signatures may span multiple lines and visibility qualifiers such as ``\texttt{static}'' may appear on a preceding line.
The LLM is instructed to return any one of:
\begin{itemize}[itemsep=1pt, topsep=3pt, leftmargin=15pt]
\item the file containing the comment-referenced code entity,
\item \texttt{NO\_MATCHED\_DEFINITION} if no suitable candidate is found,
\item \texttt{NOT\_A\_FUNCTION} if the symbol does not refer to a function.
\end{itemize}
For functions whose implementations fall outside our \texttt{*.c}/\texttt{*.h} scope (e.g., assembly implementations), the LLM is instructed to select the declaration file closest in path to the comment's file.
To guard against hallucination, we verify whether the returned file path appears in the original candidate list.
If not, we retry up to two times, each time providing the LLM with its previous erroneous output as feedback, as prior work~\cite{liu2024refining} has shown that LLM performance improves in early feedback rounds and quickly stabilizes thereafter.
If the response remains invalid after all retries, the case is marked as ``\emph{Tracing Inconclusive}'' and excluded from further tracing.
The complete prompt is available in our replication package.

After entity selection, a symbol that is confirmed to refer to a specific code entity that existed in the past---but no longer exists in the current snapshot---is regarded as a \emph{stale function reference}.
This yields the initial state $\langle \textit{sym}_0,\ \textit{file}_0,\ \textit{commit}_0 \rangle$, from which we trace the code entity's subsequent evolution.

\subsubsection{Tracking the Evolution of the Referenced Entity}
\label{sec:tracking}
Starting from the initial state $\langle \textit{sym}_0,\ \textit{file}_0,\ \textit{commit}_0 \rangle$, we iteratively trace the code entity's evolution forward along the Git history   (outlined in Algorithm~\ref{alg:trace}).
% The overall procedure is outlined in Algorithm~\ref{alg:trace}.
At each step, we locate the commit that causes the entity to disappear from its file, and pass the commit message and diff to an LLM to determine whether the entity was \emph{refactored} (i.e., its
functionality was taken over by one or more successor entities) or \emph{removed} (i.e., its functionality was eliminated entirely).
If removed, the trace terminates.
% Otherwise, we check whether the successor entity remains present in the current snapshot;
% if so, the trace is complete.
Otherwise, we check whether the successor entity still exists in the current snapshot (i.e., at \texttt{HEAD}); if so, the evolution has reached the present and the trace terminates (marked ``\emph{Aligned with HEAD}'' in Algorithm~\ref{alg:trace}).
% If not, we apply the same procedure recursively to the successor, up to a maximum depth of five iterations as a safeguard against unbounded tracing.
If not, we apply the same procedure recursively to the successor, up to a maximum depth of five iterations to guard against overly long or drifting traces.
We detail the core steps of the iteration below.

\renewcommand{\algorithmicrequire}{\textbf{Input:}}
\renewcommand{\algorithmicensure}{\textbf{Output:}}
\begin{algorithm}[t]
\scriptsize
\caption{Tracking the Evolution of a Referenced Entity}
\label{alg:trace}
\begin{algorithmic}[1]
\Require \textit{sym}: target symbol; $\textit{file}_0$: definition file at $\textit{commit}_0$;
\Statex \hspace{2.3em} $\textit{commit}_0$: symbol-introducing commit; \textit{depth}: current depth
\Ensure Evolution tree with leaf status (\emph{``Permanently Removed''}, \Statex \hspace{2.3em}
\emph{``Aligned with HEAD''}, or \emph{``Depth Limit Reached''})
\Function{trace}{\textit{sym}, \textit{file}, \textit{commit}, \textit{depth}}
    \If{$\textit{depth} \geq 5$}\ \State mark \textbf{Depth Limit Reached};\ \Return \EndIf
    \State $\textit{commit}^{rm} \gets \text{findRemovalCommit}(\textit{sym},\ \textit{file},\ \textit{commit})$
    \State $\textit{rdiff} \gets \text{filterDiff}(\text{gitShow}(\textit{commit}^{rm}),\ \textit{sym})$
    \State $\textit{result} \gets \text{validateWithRetry}(\text{llmAnalyze}(\textit{sym},\ \textit{rdiff}),\ \textit{commit}^{rm})$
    \State $\text{recordHistory}(\textit{commit}^{rm},\ \textit{result})$
    \If{$\textit{result.type} = \text{removed}$}
        \State mark \textbf{Permanently Removed};\ \Return
    \ElsIf{$\textit{result.type} = \text{refactored}$}
        \For{$s \in \textit{result.successors}$}
            \If{$\textsc{Exists}(s.\textit{sym},\ s.\textit{file},\ \text{HEAD})$}\ \State mark \textbf{Aligned with HEAD};\ \Return
            \Else\ \State \Call{trace}{$s.\textit{sym},\ s.\textit{file},\ \textit{commit}^{rm},\ \textit{depth}+1$}
            \EndIf
        \EndFor
    \EndIf
\EndFunction
\State \Call{trace}{$\textit{sym},\ \textit{file}_0,\ \textit{commit}_0,\ 0$}
\end{algorithmic}
\end{algorithm}

\textbf{1) Locating the Function-Removing Commit.}
$\mathit{commit}^{rm}_i$ is used to represent the \emph{function-removing commit} for state $\langle \mathit{sym}_i,\ \textit{file}_i,$ $\textit{commit}_i \rangle$,
i.e. the first commit after $\textit{commit}_i$ at which $\textit{sym}_i$ is no longer present in $\textit{file}_i$, regardless of whether it was deleted, renamed, or moved elsewhere.
To locate $\textit{commit}^{rm}_i$, we chronologically traverse the \emph{modification history} of $\textit{sym}_i$ on $\textit{file}_i$, i.e., all commits from $\textit{commit}_i$ to \texttt{HEAD} that modified $\textit{file}_i$ involving $\textit{sym}_i$.
The earliest commit \textit{c} at which $\lnot\textsc{Exists}(\textit{sym}_i,\ \textit{file}_i,\ c)$ is designated as $\textit{commit}^{rm}_i$.
If no such commit is found (e.g., Coccinelle misidentifies the symbol as unresolved while the function still exists, leaving no removal to locate), the case also ends in ``\emph{Tracing Inconclusive}''.

\textbf{2) LLM-Based Change Analysis.}
We provide  $\textit{commit}^{rm}_i$---its title, message, and code diff---to the LLM to determine whether $\textit{sym}_i$ was \emph{removed} or \emph{refactored}.
In the refactored case, the LLM additionally reports the successor symbols and the corresponding containing files.
The LLM can also return \emph{None}, indicating that $\textit{sym}_i$ was not actually removed at this commit (e.g., Coccinelle incorrectly identified the removing commit).
In this case, we distrust the Coccinelle-based localization and fall back to an LLM-based change analysis: we sequentially feed each commit in the modification history to the LLM, adopting the first non-\emph{None} response as the true removing commit and its change-analysis result.
If no valid removing commit can be identified, the case ends in ``\emph{Tracing Inconclusive}''.
When interacting with the LLM, we first provide the complete diff;
if the full prompt exceeds the token limit, we retain only hunks from files whose token sets shared at least one sub-token of $\textit{sym}_i$ (split on ``\_''), as renames in the Linux kernel typically preserve one or more sub-tokens of the original name.

\textbf{3) Validation and Retry.}
We validate each reported successor in two steps:
(1)~\emph{file existence check}: we verify via \texttt{git} that the reported file exists at $\textit{commit}^{rm}_i$; (2)~\emph{entity membership check}: we verify $\textsc{Exists}(s.\textit{sym},\ s.\textit{file},\ \textit{commit}^{rm}_i)$.
Accumulated error messages are fed back to the LLM, which is allowed up to two retries.
If validation still fails, this case ends in ``\emph{Tracing Inconclusive}''.
If the change is classified as \emph{removed}, the branch terminates---the function was permanently deleted at this commit.
If classified as \emph{refactored}, we check each validated successor \textit{s} against the current snapshot via \textsc{Exists}.
If \textit{s} exists, this branch has reached the present, and tracing stops.
Otherwise, \textit{s} itself was removed in a later commit, so we update the state to
\[
\langle \textit{sym}_{i+1},\ \textit{file}_{i+1},\ \textit{commit}_{i+1} \rangle
= \langle s.\textit{sym},\ s.\textit{file},\ \textit{commit}^{\rm rm}_i \rangle
\]
and continue tracing.
When a function is split into multiple successors, each successor is traced independently.

\begin{figure}[t]
    \centering
     \setlength{\abovecaptionskip}{6pt}
    \includegraphics[width=0.38\textwidth]{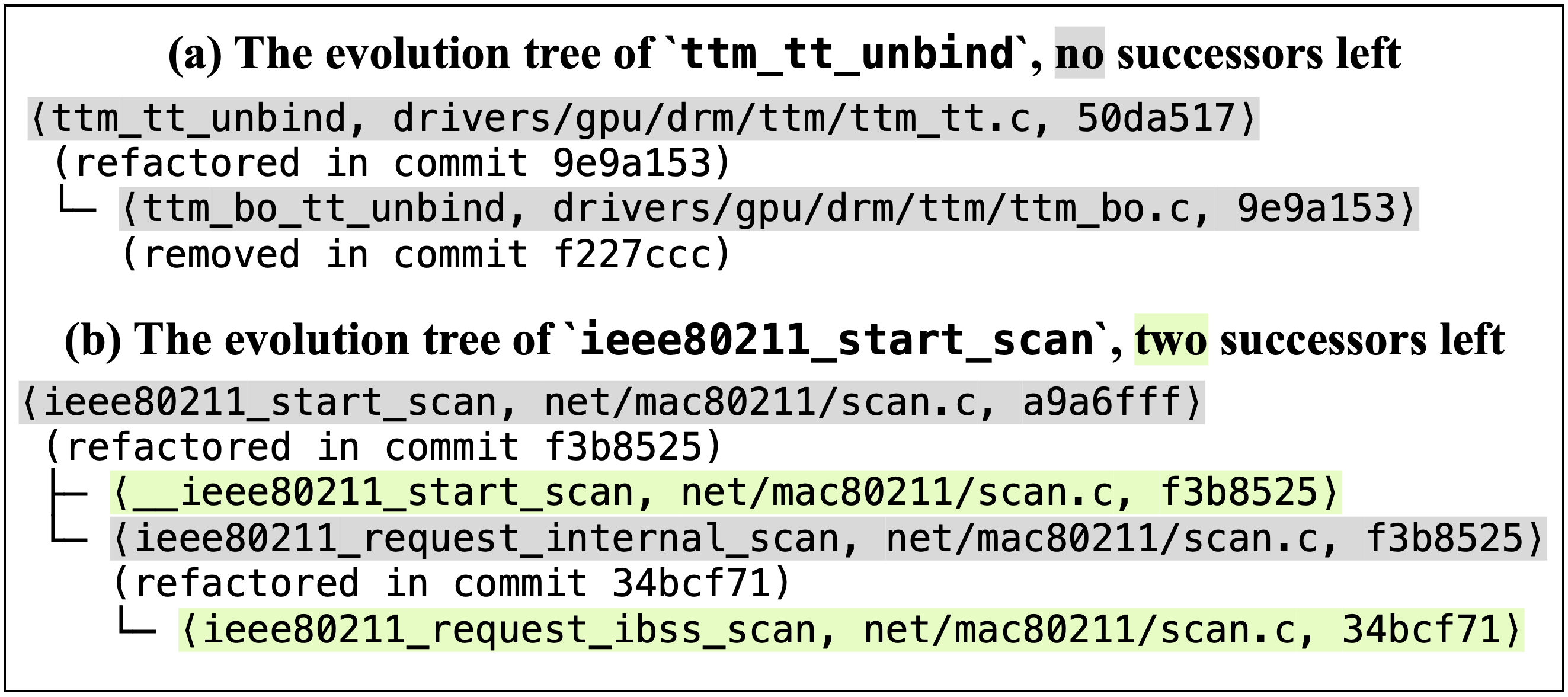}
    \caption{
    Evolution trees of two code entities. 
    \begin{revision}
    Green marks surviving successors; shaded marks those absent at \texttt{HEAD}.
    \end{revision}
    }
    \label{fig:evolution-tree}
    \vspace{-3pt}
  \end{figure}

% We organize the evolutionary history of a code entity into a tree structure.
% Figure~\ref{fig:evolution-tree} illustrates the evolution tree for two code entities.
% The function \texttt{ttm\_tt\_unbind} was relocated and renamed into \texttt{ttm\_bo\_tt\_unbind} in commit \texttt{9e9a153} titled ``drm/ttm: move ttm binding/unbinding out of ttm\_tt paths.''
% After this, \texttt{ttm\_bo\_tt\_unbind} was completely eliminated in commit \texttt{f227ccc} titled ``drm/ttm: drop unbind callback,'' without any successor.
% In another case, the function\texttt{ieee80211\_start\_scan} was split into two functions in commit \texttt{f3b8525} titled ``mac80211: fix scan races and rework scanning'':
% \texttt{\_\_ieee80211\_start\_scan}, which inherited the core scanning logic and still exists in the current version, marking the end of the trace;
% and \texttt{ieee80211\_request\_internal\_scan}, which was later refactored in commit \texttt{34bc-f71} titled ``mac80211: fix ibss scanning'' and renamed to \texttt{ieee80211\_request\_ibss\_scan}, a state that has been maintained until today.
We organize the evolutionary history of a code entity into a tree structure.
Figure~\ref{fig:evolution-tree} illustrates two examples.
The function \texttt{ttm\_tt\_unbind} was first renamed to \texttt{ttm\_bo\_tt\_unbind} and later removed entirely, leaving no surviving successor.
In contrast, \texttt{ieee80211\\\_start\_scan} was split into two functions: \texttt{\_\_ieee80211\_start\_scan}, which inherited the core scanning logic and still exists today, and \texttt{ieee80211\_request\_internal\_scan}, which was further renamed to \texttt{ieee80211\_request\_ibss\_scan}.

\subsection{Generating Repair Suggestions with LLM}
\label{sec:repair}

To repair stale references in comments, we generate suggestions based on the evolution history of the referenced code entities.
Each stale reference is treated individually.
The prompt is organized into four input sections followed by a chain-of-thought analysis request.
(The full prompt details are included in our release package.)

The structured input consists of:
\textbf{1) Target Comment Context}, which includes the current comment, its file location, the specific stale reference, and the surrounding code context in the current version, i.e., the environment into which the repaired comment needs to fit;
\textbf{2) Origin Context}, which includes the description of the symbol-introducing commit, together with the comment's code context and the original function's source at that time, allowing the LLM to contrast the original and current versions and judge whether the comment's intent still holds;
\textbf{3) Evolutionary History}, which includes the full evolution tree of the referenced function (in a format similar to Figure~\ref{fig:evolution-tree}), followed by detailed change logs, including each function-removing commit's title and message and a pruned diff restricted to the disappeared entity or its successors;
\textbf{4) Evolution Outcomes}, where we present the current code of surviving successors in the evolution tree, so the LLM can assess whether they are suitable replacements for the stale reference.

We then instruct the LLM to reason as follows before answering:
\textbf{1) Evolution Analysis}: trace the evolution path of the missing function---whether it was simply renamed, had its logic split, or was entirely removed---and identify which surviving function, if any, carries the semantic equivalent of the original reference.
\textbf{2) Comment Context \& Semantic Analysis}: examine the current code context surrounding the comment and assess whether the comment's original intent remains valid given the code changes.
\textbf{3) Repair Strategy}: propose a concrete fix by updating the function name, rephrasing the comment when needed, or entirely removing it if it has become obsolete, ensuring the repaired comment fits the current code context.

Finally, the LLM produces the repaired comment based on the above analysis.
For example, for the comment ``\textit{See the link below and \textbf{ieee80211\_start\_scan()} for more.}'', \textsc{ReCite} recommends replacing the stale reference with its core-logic successor \textit{\_\_ieee80211\_start\\\_scan} (Figure~\ref{fig:evolution-tree}).
This repair has been accepted into the Linux kernel.
% \footnote{\url{https://git.kernel.org/pub/scm/linux/kernel/git/netdev/net-next.git/commit/?id=b8a57b979a7c2069081ed0bf88a727b17d85ac96}}

\section{Experimental Setup}
\label{sec:setup}

\begin{figure}[t]
    \centering
    \setlength{\abovecaptionskip}{6pt}
    \includegraphics[width=0.40\textwidth]{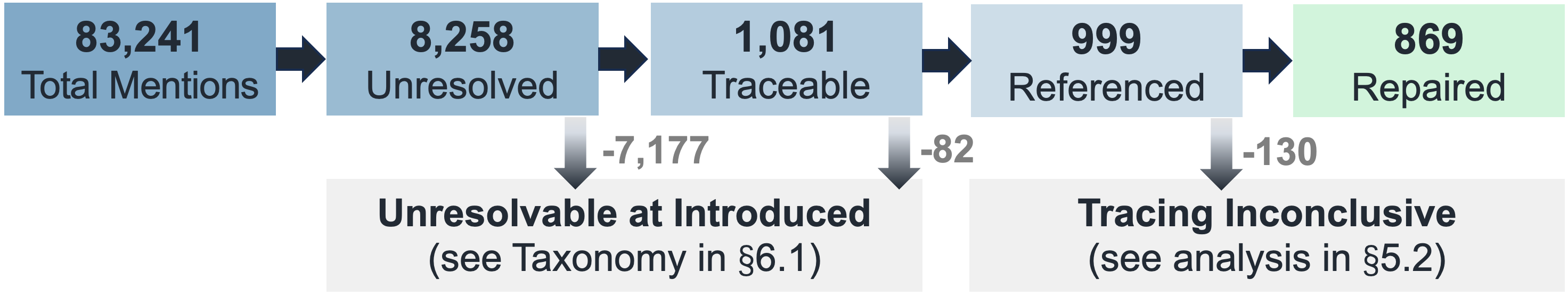}
    \caption{
    \begin{revision}
    Attrition flow of \textsc{ReCite}'s stale-function-reference detection and repair pipeline.
    \end{revision}
    }
    \label{fig:flow_diagram}
\end{figure}

%\begin{revision}
We evaluated \textsc{ReCite} on Linux kernel v6.18-rc1 for automatically identifying and repairing stale function references in kernel comments.
From this snapshot, we extracted \verified{49,302} unique function-form symbols across \verified{61,866} comments, totaling \verified{83,241} mentions.
Using grep with Coccinelle, \textsc{ReCite} flags 8,258 function-form mentions (5,442 unique symbols across 7,111 comments) as unresolved in the current snapshot.
Among them, we found 7,177/8,258 (86.9\%) mentions were already unresolvable when the symbol was first introduced into the comment; we provide a manual categorization of their underlying reasons in Section~\ref{sec:taxonomy} (only 17.3\% of these are genuinely missing functions).
For the remaining 1,081/8,258 (13.1\%) resolvable mentions, for which at least one same-name candidate can serve as a traceable starting point, \textsc{ReCite} proceeds to construct the evolution history:
it first locates a valid referenced code entity for 999 of them (the other 82 lack a valid entity),
then successfully traces and repairs 869 of these (the remaining 130 end in \emph{Tracing Inconclusive}).
\textbf{These 869 cases therefore constitute \textsc{ReCite}'s positive predictions for repair-worthy stale function references.}
Our manual analysis shows that the majority of these \verified{130} cases are upstream Coccinelle false positives that in fact need no repair (see Section~\ref{sec:resultsRQ2});
accordingly, \textsc{ReCite} does not treat them as repair-worthy stale references and does not present repair suggestions to the user.
The 82 cases without a valid comment-referenced code entity are effectively unresolvable at introduction (the static-analysis candidates are only same-name objects); their underlying reasons can also be understood through the taxonomy established in Section~\ref{sec:taxonomy}.
We summarize the above attrition flow in Figure~\ref{fig:flow_diagram}.
%\end{revision}

Throughout all LLM-based steps, we used DeepSeek-V3.2~\cite{liu2025deepseek} as a cost-efficient model with temperature set to 0 to reduce output randomness.
To assess both \textsc{ReCite}'s component-level performance (RQ1--RQ3) and its end-to-end performance (RQ4), we posed following research questions: 
% \textbf{RQ1}: How accurately can \textsc{ReCite} detect unresolved function-form symbols? 
% \textbf{RQ2}: How accurately does \textsc{ReCite} trace the function-evolution history? 
% \textbf{RQ3}: How effectively can \textsc{ReCite} repair the identified stale function references? 
% \textbf{RQ4}: How accurately does \textsc{ReCite} identify repair-worthy stale function references end to end?
\begin{itemize}[itemsep=1pt, topsep=3pt, leftmargin=15pt]
\item \textbf{RQ1}: How accurately can \textsc{ReCite} detect unresolved function-form symbols?
\item \textbf{RQ2}: How accurately does \textsc{ReCite} trace the function-evolution history? 
\item \textbf{RQ3}: How effectively can \textsc{ReCite} repair the identified stale function references? 
\item \textbf{RQ4}: How accurately does \textsc{ReCite} identify repair-worthy stale function references end to end?
\end{itemize}

\subsection{RQ1: Detection of Unresolved Symbols}
As described in Section~\ref{sec:identification}, \textsc{ReCite} uses a two-step process to determine whether a symbol is resolved at a given version: a coarse-grained search with \texttt{grep} followed by fine-grained static analysis with Coccinelle.
To evaluate the accuracy of this pipeline, we replaced Coccinelle with three other widely used static analysis tools: \textbf{Tree-sitter}~\cite{tree-sitter} (a concrete syntax tree parser), \textbf{Universal Ctags}~\cite{universal-ctags} (a syntax-based symbol indexer), and \textbf{GNU Global}~\cite{gnu-global} (a project-wide cross-reference tool).
Each was configured to check whether a matching function definition, declaration, or function-like macro definition exists.
We also evaluated using \texttt{grep} alone without any fine-grained analysis.
We denote the five resulting configurations as
\textsc{Resolve}-Coccinelle (grep + Coccinelle, our default),
\textsc{Resolve}-TreeSitter (grep + Tree-sitter),
\textsc{Resolve}-Ctags (grep + Universal Ctags),
\textsc{Resolve}-GNUGlobal (grep + GNU Global),
and \textsc{Resolve}-OnlyGrep (grep only, without any fine-grained static analysis).
We ran all five configurations on the 49,302 unique function-form symbols extracted from \texttt{HEAD}.
Among the \verified{49,302} symbols, \verified{3,796} received conflicting resolution results across the five configurations.
% Table~\ref{tab:unresolved-overview} shows the distribution of unresolved symbols across tools.
To establish ground truth, we randomly sampled 200 of these disagreement cases for manual annotation, forming the \textbf{Symbol Resolution 200} dataset.
Two annotators---both senior undergraduate students majoring in software engineering---independently labeled each symbol as \emph{resolved} or \emph{unresolved} at \texttt{HEAD}, by inspecting the source code at the target commit.
They first annotated a pilot batch of 20 symbols to calibrate their criteria, achieving a Cohen's $\kappa$ of \verified{0.89}.
After resolving disagreements through discussion with the first author, the remaining symbols were split evenly between the two annotators.
\begin{revision}
In the final ground truth, \verified{110} symbols were labeled as \emph{resolved} and \verified{90} as \emph{unresolved}.
\end{revision}
We evaluated all five \textsc{Resolve} variants on this dataset and report precision ($P$), recall ($R$), F1-score ($F1$), and accuracy ($Acc$).

\subsection{RQ2: Accuracy of Evolution Tracing}
% From Linux kernel v6.18-rc1, we extracted \verified{49,302} unique function-form symbols across \verified{61,866} comments, totaling \verified{83,241} mentions.
% Using $\textsc{Resolve}$-Coccinelle, we detected \verified{5,442} \textit{unresolved} symbols across \verified{7,111} comments, totaling \verified{8,258} mentions.
% When constructing the evolution history, we found that \verified{7,177/8,258 (86.9\%)} mentions were already unresolvable when the symbol was first introduced into the comment, lacking a traceable starting point; we provide a manual categorization of their underlying reasons in Section~\ref{sec:taxonomy}.
% For the remaining \verified{1,081 (13.1\%)} mentions---where the symbol could be resolved at its symbol-introducing commit---\textsc{ReCite} constructs the function evolution history for repair.
% We evaluate two sub-tasks to assess the quality of the constructed evolution history:
For the \verified{1,081} mentions resolvable at their symbol-introducing commit, \textsc{ReCite} constructs the evolution history for repair.
We evaluate this stage through two sub-tasks:
(1)~\emph{comment-reference entity selection}---whether \textsc{ReCite} correctly identifies the referenced code entity; and
(2)~\emph{refactoring analysis}---whether \textsc{ReCite} correctly determines how the entity was changed at function-removing commits.

\vspace{0.5em}
\noindent\textbf{Comment-Reference Entity Selection.}%
\label{sec:eval-entity-selection}
\begin{revision}
We randomly sampled 200 cases from the \verified{1,081} resolvable mentions to form the \textbf{Entity Selection 200} dataset.
\end{revision}
Two annotators independently labeled each case, determining:
(a)~whether the symbol refers to a concrete function entity, and if so,
(b)~which file contains the referenced entity at $\textit{commit}_0$.
The annotator selected one of:
a file path (the entity is found),
\texttt{NOT\_A\_FUNCTION} (the symbol does not refer to a function), or
\texttt{NO\_MATCHED\_DEFINITION} (no suitable candidate is found).
The annotators first labeled a pilot batch of 20 cases.
At the binary level (entity exists vs.\ not), they achieved a Cohen's $\kappa$ of \verified{0.62} (substantial agreement).
Among the 16 cases where both identified a valid entity, they agreed on the exact file path in all cases (100\%).
After resolving disagreements through discussion with the first author, the remaining cases were split evenly between the two annotators.
\begin{revision}
In the final ground truth, 173 symbols were labeled as having a valid code entity, while 27 were not (14~\texttt{NO\_MATCHED\_DEFINITION}, 13~\texttt{NOT\_A\_FUNCTION}).
\end{revision}

We also evaluated \textsc{ReCite} on two levels.
First, we assessed \emph{entity detection}: whether \textsc{ReCite} correctly determines that a symbol has a valid code entity.
We mapped the output to a binary label: $y{=}1$ if \textsc{ReCite} returns a file path; $y{=}0$ if it returns \texttt{NOT\_A\_FUNCTION} or \texttt{NO\_MATCHED\_DEFINITION}.
We then report $P$, $R$, and $F1$.
Second, we assessed \emph{entity localization}: among the cases where both \textsc{ReCite} and the ground truth agree that a valid entity exists ($y{=}1$), we computed the accuracy of the returned file path, i.e., the proportion of cases where \textsc{ReCite}'s file path matches the annotator's file path.

\vspace{0.5em}
\noindent\textbf{Refactoring Analysis.}%
\label{sec:eval-refactoring}
Applying entity selection to \verified{1,081} resolvable mentions, \textsc{ReCite} located a valid referenced code entity for \verified{999} of them.
To evaluate how accurately \textsc{ReCite} traces their evolution, we use RefDiff~\cite{silva2020refdiff}, a static-analysis-based refactoring detection tool, as a baseline.
Using \textsc{ReCite}, we successfully constructed evolution trees for \verified{869} of \verified{999} candidate mentions, containing a total of \verified{916} unique function-removing commits.
The remaining \verified{130} cases ended in \emph{Tracing Inconclusive}, analyzed in Section~\ref{sec:resultsRQ2}.
Using RefDiff, we constructed evolution trees for 903 cases, containing 837 function-removing commits.
The two approaches shared \verified{785} removing commits where they analyzed the same code entity $\langle \textit{sym}, \textit{file} \rangle$ at the same commit, of which \verified{282} yielded different results.
% and \verified{503} yielded the same result.

\begin{revision}
% ### Disproportionate Stratified Sampling Version：
% Following the RQ1 setup, we used disproportionate stratified sampling:
% we randomly sampled 100 disagreement cases and 50 agreement cases for manual annotation, forming the \textbf{Refactoring Analysis 150} dataset.
% ### Original Version：
We randomly sampled 200 cases from these divergent cases to form the \textbf{Refactoring Analysis 200} dataset.
\end{revision}
For each case---a code entity $\langle \textit{sym}, \textit{file} \rangle$ at a specific function-removing commit---the annotator determined whether the entity was \emph{removed} (deleted without a successor) or \emph{refactored}.
If refactored, the annotator identified each successor as a \emph{refactoring route}:
$\langle \textit{sym}_A, \textit{file}_A \rangle \rightarrow \langle \textit{sym}_B, \textit{file}_B \rangle$,
where $\langle \textit{sym}_A, \textit{file}_A \rangle$ is the entity before the commit and $\langle \textit{sym}_B, \textit{file}_B \rangle$ is its successor after the commit.
A single entity may yield multiple refactoring routes when the function is split into several successors.

Two annotators independently labeled a pilot batch of 20 cases.
They agreed on 19 (95\%) cases for the binary judgment, with nearly all labeled as \emph{refactored}.
We computed pairwise agreement by treating one annotator's routes as predictions against the other's, yielding \verified{$F1{=}0.80$}.
After resolving disagreements through discussion with the first author, they independently labeled the remaining cases.
\begin{revision}
% ### Disproportionate Stratified Sampling Version：
% In the final ground truth, the disagreement subset contains 93 \emph{refactored}, 5 \emph{removed}, and 2 Coccinelle false-positive cases (the function still exists and is therefore excluded when evaluating), while the agreement subset contains 33 \emph{refactored} and 17 \emph{removed} cases.
% ### Original Version：
In the final ground truth, 191 cases are labeled as \emph{refactored}, 7 as \emph{removed}, and 2 as Coccinelle false positive---the function still exists at the supposed removing commit and was not actually changed.
\end{revision}

We evaluated both \textsc{ReCite} and RefDiff on this dataset at two levels:
(1)~\emph{binary classification}---whether a function-removing commit is correctly classified as \emph{refactored} or \emph{removed}, with \emph{refactored} as the positive class, measured by $P$, $R$, and $F1$ (the Coccinelle false positive is excluded, leaving \rev{198} valid cases); and
(2)~\emph{successor identification}---among commits where both the tool and the ground truth agree on \emph{refactored}, whether the tool identifies the correct successor(s), measured by $P$, $R$, and $F1$ at the refactoring-route level.
\begin{revision}
% ### Disproportionate Stratified Sampling Version：
% As in RQ1, we report results separately for the disagreement subset ($n{=}100$), the agreement subset ($n{=}50$), the full Refactoring Analysis 150 sample, and a weighted overall estimate.
% ### Original Version：
% None for Original Version here.
\end{revision}
\vspace{-1em}

\subsection{RQ3: Effectiveness of Repair}
We manually evaluated the quality of LLM-generated comment repairs.
The first author reviewed an initial set of 30 cases to establish the following four-category rubric:
\begin{itemize}[itemsep=1pt, topsep=3pt, leftmargin=15pt]
  \item \textbf{Acceptable}: the repair appropriately identifies the successor or handles the removal, and can be directly applied.
  \item \textbf{Partial}: the repair makes partially correct changes but is not directly applicable, e.g., leaving residual outdated descriptions.
  \item \textbf{Incorrect}: the suggested repair is entirely wrong and offers no reference value.
\item \textbf{Off-Target}: the case falls outside the scope of our repair task, e.g., the symbol does not refer to a function.
\end{itemize}

\begin{revision}
We then randomly sampled an additional 200 cases to form the \textbf{Repair Quality 200} dataset.
\end{revision}
Two annotators (the same for the Symbol Resolution 200 dataset) independently labeled a pilot batch of 20 cases, achieving a Cohen's $\kappa$ of \verified{0.72}.
After resolving disagreements through discussion with the first author, the remaining cases were divided between two annotators.

\begin{revision}
\vspace*{0.1cm}
\noindent\textbf{Ablation Study.}
Additionally, we include two ablation studies to assess the necessity of LLM-based repair and evolution history. 
For the LLM-based repair ablation, we construct a rule-based repair baseline that, instead of an LLM, applies simple rules over the evolution tree: 
if the evolution tree contains a surviving successor, the baseline replaces the stale function mention with the successor whose name has the smallest edit distance to the original; otherwise, it removes the enclosing sentence.
For the evolution-history ablation, we implement a no-history \textsc{ReCite} variant by removing the evolution-history input from the repair stage, leaving the LLM only with the original and current comment context.
Comparing \textsc{ReCite} against these two variants isolates the contribution of LLM-based repair and evolution-history context.
\end{revision}

\begin{revision}
\subsection{RQ4: End-to-End Detection.}
Of the \verified{83,241} function-form mentions in v6.18-rc1, \textsc{ReCite} flags \verified{869} (\verified{1.0\%}) as repair-worthy stale function references and leaves the remaining \verified{82,372} (\verified{99.0\%}) unflagged.
To assess this end-to-end detection, we construct the \textbf{Repair-Worthy Detection 400} dataset via \textit{disproportionate stratified sampling}:
200 predicted-positive cases reused from the Repair Quality 200,
and 200 predicted-negative cases randomly sampled from the remaining \verified{82,372} unflagged mentions.
Manual annotation of the predicted-positive sample, as reported in RQ3, confirms 178 true positives and 22 false positives.
We further annotate the 200 predicted-negative cases (the two annotators agreed on all 20 pilot cases), and find only 2 genuinely stale function references that require repair.
As the closest applicable baseline, we include C4RLLaMA~\cite{rong2025code}, a learning-based comment--code inconsistency detector.
We reproduce it under its original setting and adapt it to our task by taking its \emph{inconsistent} verdict as a positive flag.
We report the performance of \textsc{ReCite} and C4RLLaMA on the predicted-positive ($n{=}200$), the predicted-negative ($n{=}200$), and the full balanced ($n{=}400$) samples, using $P$, $R$, $F1$, and $Acc$.
% To estimate population-level performance, we additionally report a weighted overall result, where the two strata are weighted by their prevalence in the full population: 869 flagged mentions and 82,372 unflagged mentions.
% We summarize the five annotated datasets used for evaluation in Table~\ref{tab:evaluation-overview}.
\end{revision}

\begin{table}[t]
\centering
\scriptsize
\caption{
\begin{revision}
(RQ1) \verified{Performance of five $\textsc{Resolve}$ variants on the \textbf{Symbol Resolution 200} dataset. \emph{Resolved} is the positive class.}
\end{revision}
}
\label{tab:rq1-results}
\vspace{-3pt}
\begin{tabular}{ccccc}
\toprule
\textbf{Variant} & \textbf{$P$} & \textbf{$R$} & \textbf{$F1$} & \textbf{$Acc$} \\
\midrule
$\textsc{Resolve}$-Coccinelle  & \textbf{96.7\%} & 79.1\% & \textbf{0.870} & \textbf{87.0\%} \\
$\textsc{Resolve}$-TreeSitter & 95.4\% & 56.4\% & 0.709 & 74.5\% \\
$\textsc{Resolve}$-Ctags      & 50.0\% & 53.6\% & 0.518 & 45.0\% \\
$\textsc{Resolve}$-GNUGlobal      & 75.4\% & 41.8\% & 0.538 & 60.5\% \\
$\textsc{Resolve}$-OnlyGrep   & 55.0\% & \textbf{100.0\%} & 0.710 & 55.0\% \\
\bottomrule
\end{tabular}
\end{table}

\section{Results}
\label{sec:results}

\subsection{RQ1: Detection of Unresolved Symbols}

\begin{revision}
% ### Disproportionate Stratified Sampling Version：
% Table~\ref{tab:rq1-results} reports the performance of five $\textsc{Resolve}$ variants on the \textbf{Symbol Resolution 150} dataset.
% On the agreement subset, all five variants achieve 100\% accuracy.
% On the disagreement subset, $\textsc{Resolve}$-Coccinelle achieves the highest $F1$ (\verified{0.864}) and accuracy (\verified{83.0\%}) among all variants, which also gives it the best weighted overall performance (\verified{$F1{=}0.993$}, \verified{$Acc{=}98.7\%$}).
% ### Original Version：
% Given the comparatively lower recall of $\textsc{Resolve}$-Coccinelle, we analyzed 23 false-negative cases---symbols whose corresponding functions exist but were missed by it.
% The primary cause is \emph{macro-wrapped or macro-generated function definitions}, whose definitions are syntactically hidden inside C preprocessor macros.
% For instance, \texttt{PAGE\_TYPE\_OPS(Buddy, ...)} constructs the function named \texttt{PageBuddy} through token pasting (\texttt{\#\#}), making it unrecognizable as a function definition at the source level.
% Because our SmPL rules operate on unexpanded source text, such patterns cannot be matched by the current analysis.
% This is a deliberate trade-off in our tool configuration, which we discuss further in Section~\ref{sec:threats}.

% We find that Coccinelle recognizes common Linux-kernel macro forms (e.g., \texttt{BPF_CALL}) through built-in macro hints.
Table~\ref{tab:rq1-results} reports the performance of five $\textsc{Resolve}$ variants on the \textbf{Symbol Resolution 200} dataset.
$\textsc{Resolve}$-Coccinelle achieves the highest $F1$ (\verified{0.870}) and accuracy (\verified{87.0\%}) among all variants.
\end{revision}
We find that Coccinelle can recognize common Linux-kernel macro forms that match its predefined patterns (e.g., \texttt{BPF\_CALL}).
For example, for \texttt{BPF\_CALL\_2(bpf\_redirect, ...)}, only $\textsc{Resolve}$-Coccinelle and $\textsc{Resolve}$-OnlyGrep (by string matching) classify \texttt{bpf\_redirect} as resolved, while the other three tools all classify it as unresolved.
However, it remains limited in handling \emph{macro-generated function definitions}, whose definitions are syntactically hidden inside C preprocessor macros, which accounts for its comparatively lower recall ($R{=}79.1\%$). 
For instance, \texttt{PAGE\_TYPE\_OPS(Buddy, ...)} constructs the function named \texttt{PageBuddy} through token pasting (\texttt{\#\#}), making it unrecognizable as a function definition at the source level (grep survives as the name appears elsewhere).
Because our SmPL rules operate on unexpanded source text, such patterns cannot be matched by the current analysis.
This is a deliberate trade-off in our tool configuration, which we discuss further in Section~\ref{sec:threats}.\looseness=-1

\begin{table}[t]
\centering
\scriptsize
\setlength{\belowcaptionskip}{2pt}
\caption{
\begin{revision}
(RQ2) Performance on \textbf{Entity Selection 200} dataset.
\end{revision}
}
\label{tab:entity-selection}
\vspace{-3pt}
\begin{tabular}{cl}
\toprule
\textbf{Task} & \multicolumn{1}{c}{\textbf{Metric}} \\
\midrule
\multirow{2}{*}{\textbf{\textit{Entity detection}}} & $\mathit{TP}=173$, \enspace $\mathit{FP}=13$, \enspace $\mathit{FN}=0$, \enspace $\mathit{TN}=14$ \\
 & $P = 93.0\%$ \quad $R = 100.0\%$ \quad $F1 = 0.964$ \\
\midrule
\textbf{\textit{File path match}} & On 173 TP cases, $Acc = 98.3\%$ \\
\bottomrule
\end{tabular}
\end{table}

\subsection{RQ2: Accuracy of Evolution Tracing}
\label{sec:resultsRQ2}

\subsubsection{Comment-Reference Entity Selection}
\label{sec:entity-selection-results}

\begin{revision}
As shown in Table~\ref{tab:entity-selection},
\textsc{ReCite} achieves an $F1$ of 0.964 on entity detection.
The 13 false positives indicate that some symbols without a valid code entity may enter the subsequent tracing and repair steps, but this conservative strategy ensures perfect recall ($R{=}100\%$, $\mathit{FN}{=}0$), so no symbol that genuinely requires repair is missed.
Among the 173 true-positive cases, \textsc{ReCite} returns the correct file path in 170 (98.3\%).
These results indicate that our comment-reference entity selection provides a reliable starting point for evolution tracing and repair.
\end{revision}

% Among the remaining same-file cases, the median distance between the comment and the referenced function is \verified{92} lines.
% with a median same-file distance of \verified{304} lines.
We additionally analyze the relative positions between the referenced entities and the comments that mention them.
Among the 173 annotated true-positive cases, 55.5\% involve a cross-file reference, i.e., the entity definition resides in a different file from the comment.
A similar proportion can be observed at larger scale: of the 999 valid entities \textsc{ReCite} identifies from all 1,081 resolvable mentions, 60.5\% are cross-file references.
These findings indicate that cross-file entity references are prevalent in Linux kernel comments, and the simple local-context analysis is insufficient to address them, validating \textsc{ReCite}'s LLM-based cross-file entity resolution.

\subsubsection{Refactoring Analysis.}
\begin{revision}
As shown in Table~\ref{tab:refactoring-analysis}, \textsc{ReCite} substantially outperforms RefDiff in both distinguishing whether a function was refactored or removed \verified{($F1$: 0.974 vs.\ 0.460)} and identifying correct successors \verified{($F1$: 0.812 vs.\ 0.577)}.
\end{revision}
\textsc{ReCite} can benefit from the semantic information in commit messages~\cite{lyu2026agentszz}.
For example, commit \texttt{5e7a5c8} is titled ``\textit{fold dentry\_kill() into dput()}''.
\textsc{ReCite} correctly identifies \texttt{dput()} as the successor---an Inline Function refactoring where \texttt{dentry\_kill()}'s logic is absorbed into an existing, larger function---while RefDiff only classifies this case as \emph{removed}.
We find that \textsc{ReCite} shows a relatively lower precision on successor identification ($P{=}0.736$), as it sometimes reports semantically paired functions as successors, e.g., for the renamed \texttt{set\_settings()}, it returns the getter \texttt{get\_link\_ksettings()} alongside the true successor \texttt{set\_link\_ksettings()}.
Regarding false negatives, it occasionally misses semantically distant successors:
for \texttt{vmtruncate()}, it identifies the core successor \texttt{truncate\_pagecache()} but misses \texttt{inode\_newsize\_ok()}, an auxiliary validation helper extracted from it.
Nevertheless, \textsc{ReCite}'s high recall ($R{=}0.905$) is valuable for downstream repair, ensuring potential successors are not missed.

\begin{table}[t]
\centering
\scriptsize
\caption{
% \begin{revision}
(RQ2) \verified{Performance on the \textbf{Refactoring Analysis 200} dataset.}
Binary classification: \emph{refactored} is the positive class ($n{=}198$, 2 Coccinelle false positives excluded).
Successor identification: $n$ = cases where tool and human agree on \emph{refactored}.
% \end{revision}
}
\label{tab:refactoring-analysis}
\vspace{-3pt}
\begin{tabular}{llcccc}
\toprule
\textbf{Task} & \textbf{Tool} & $n$ & $P$ & $R$ & $F1$ \\
\midrule
\multirow{2}{*}{Binary classification}
 & \textsc{ReCite} & 198 & \textbf{0.974} & \textbf{0.974} & \textbf{0.974} \\
 & RefDiff & 198 & 0.951 & 0.304 & 0.460 \\
\midrule
\multirow{2}{*}{Successor identification}
 & \textsc{ReCite} & 186 & \textbf{0.736} & \textbf{0.905} & \textbf{0.812} \\
 & RefDiff & 58 & 0.672 & 0.506 & 0.577 \\
\bottomrule
\end{tabular}
\end{table}

Across all 869 successfully constructed evolution trees, \textsc{ReCite} identifies a total of \verified{1,136} function-removing commits, averaging 1.31 per tree.
The deepest trees reach depth 6.
For example, the function \texttt{ctl\_clear\_bit()} undergoes 5 rounds of refactoring, but one of its leaf nodes \texttt{system\_ctl\_set\_clear\_bit()} is absent from the current codebase after reaching the iteration limit, so this case is marked as \emph{Tracing Inconclusive}.
However, its other leaf node \texttt{system\_ctl\_clear\_bit()} is present in the current codebase and can serve as a valid repair target.
Another case, \texttt{estimation\_timer()}, illustrates the breadth of a single evolution: in commit \texttt{705dd344}, it is refactored into 6 successors---1 for top-level execution loop, 2 for core estimation logic, and 3 for lifecycle management.
These cases highlight the complexity of kernel code evolution, motivating the need for automated tracing support.
Of the 869 trees, \verified{335 (38.6\%)} end with no surviving successor.
The remaining \verified{534 (61.4\%)} retain at least one successor in the current codebase, providing concrete targets for repairing the stale comment.

We also analyze the 130 out of 999 cases (13.0\%) marked as \emph{Tracing Inconclusive}, where the evolution tree construction does not reach a definitive conclusion.
One of them is the \texttt{ctl\_clear\_bit()} case described above.
\verified{87} of these (67\%) arise because no removing commit is found during the traversal of the modification history.
We sample ten such cases and confirm that the function still exists at \texttt{HEAD}, meaning the upstream Coccinelle check incorrectly flagged it as unresolved at the current version, so no repair is actually needed.
The remaining \verified{42} cases (32\%) fail during successor validation after all retries.
We manually inspect ten of these and find that seven are caused by an incorrectly located removing commit leading the LLM to analyze the wrong diff.
The other three cases represent corner cases not yet covered, such as a non-function successor (e.g., a function replaced by a \texttt{struct} that cannot pass the function-level existence check).
Overall, ``\emph{Tracing Inconclusive}'' filters cases requiring no repair or affected by localization errors, leaving 869 reliable evolution trees for repair.

% Overall, the ``\emph{Tracing Inconclusive}'' mechanism serves as a safety net:
% most filtered cases either require no repair or stem from intermediate localization errors, 
% ensuring that the 869 evolution trees proceeding to repair are built on reliable results.

\subsection{RQ3: Effectiveness of Repair}

Among 200 annotated \textsc{ReCite}'s repair suggestions, 85 (42.5\%) are labeled \textbf{Acceptable} (directly applicable), 93 (46.5\%) \textbf{Partial} (providing useful reference for manual repair), 0 (0.0\%) \textbf{Incorrect}, and 22 (11.0\%) \textbf{Off-Target} (outside our repair scope).
% No repair is judged entirely wrong or misleading.
We further analyze the sub-patterns within each category.

% \vspace{0.5em}
\noindent\textbf{Acceptable (85).}

\begin{revision}
\noindent\textbf{1) Proper Name Substitution (71/85)}:
\end{revision}
\textsc{ReCite} locates a suitable successor, and directly substituting the function name suffices to fix the stale reference.

\begin{revision}
\noindent\textbf{2) Proper Wording Revision (11/85)}:
\end{revision}
\textsc{ReCite} goes beyond simple name replacement and also updates other outdated descriptions in the comment.
For example, the comment \textit{``it will be used as xxx\_of\_node on \textbf{soc\_bind\_dai\_link()}''}
references a function, whose successor is \texttt{snd\_soc\_add\_pcm\_runtime()}.
Since \texttt{snd\_soc\_add\newline\_pcm\_runtime()} no longer directly holds the \texttt{xxx\_of\_node} structure but uses it for component matching, \textsc{ReCite} revises the comment to \textit{``xxx\_of\_node for component matching in \textbf{snd\_soc\_add\_pcm} \textbf{\_runtime()}''} to better reflect this change.

\begin{revision}
\noindent\textbf{3) Proper Sentence Deletion (3/85)}:
\end{revision}
\textsc{ReCite} correctly suggests deleting the clause that mentions a removed function, such as \texttt{irda\_connect()} in \textit{``Copied and modified \textbf{irda\_connect()}''}.

We also find that \textsc{ReCite} exhibits a degree of tolerance for intermediate errors.
For example, the evolution tree for \texttt{swiotlb\_unmap\\\_page()} traces an incorrect path, yet the LLM bypasses it and identifies the correct replacement \texttt{xen\_swiotlb\_unmap\_page()} from the current code context, still producing an Acceptable repair.

\vspace{0.5em}
\noindent\textbf{Partial (93).}
Complex code evolution can also cause \textsc{ReCite}'s repair to be incomplete, in two ways.
First, \textsc{ReCite} may correctly identify a successor but leave residual outdated wording in the repaired comment (34 cases).
Second, when no single successor can fully assume the original function's role, \textsc{ReCite} instead replaces the stale name with a natural-language description (59 cases).
We find that the repairs proposing a concrete successor are more readily accepted and easier to verify; natural-language replacements, while less likely to become outdated, lose the direct code-navigation value that a concrete function name offers---we thus tend to label such repairs as Partial.
Among the 59 natural-language-description cases, 39 are labeled as having a more precise alternative available, while the remaining 20 need a more focused description.

\begin{revision}
\noindent\textbf{1) More Wording Revision Needed (34/93)}:
\end{revision}
\textsc{ReCite} correctly identifies the successor, but the repair still contains residual outdated or inaccurate descriptions.
For instance, \texttt{amdgpu\_fence\_driver\\\_init\_ring} has the comment \textit{``Helper function for \textbf{amdgpu\_fence\\\_driver\_init()}''}, \textsc{ReCite} correctly traces the rename to \texttt{amdgpu\_fence\\\_driver\_sw\_init()} and substitutes the name.
However, the body of \texttt{amdgpu\_fence\_driver\_sw\_init()} now consists solely of \texttt{return 0}---the helper relationship no longer holds, and the comment should be removed rather than merely updated.

\begin{revision}
\noindent\textbf{2) More Precise Alternative Available (39/93)}:
\end{revision}
Annotators identify a more precise alternative for these cases---33 have a suitable function name available in the codebase, and 6 warrant deletion.
For example, \texttt{pump\_transfers()} was split into three sub-functions during an SPI core migration, none of which alone preserves the original semantics.
\textsc{ReCite} substitutes \textit{``Concurrent handling of the same transfer completion leads to problems''}, but annotators find that \texttt{spi\_finalize\_current\_transfer()}---the function that now handles transfer completion in the new architecture and already appears in the collected repair context---is a more precise replacement.

\begin{revision}
\noindent\textbf{3) Too Broad Description (20/93)}:
\end{revision}
% No suitable function name exists, but the general descriptions are too broad or verbose, requiring manual adjustment.
No suitable function name exists, so using the natural-language description is an appropriate choice, but the resulting description is too broad or verbose, requiring manual adjustment.

\vspace{0.5em}
\noindent\textbf{Off-Target (22):}

\begin{revision}
\noindent\textbf{1) Changelog Comment (9/22)}:
\end{revision}
Nine cases appear in changelog-style comments (e.g., \textit{``1999/11\ldots Made \textbf{nbd\_end\_request()} use the io\_request\_lock''}), which record historical changes rather than describe current code and thus do not require repair.

\begin{revision}
\noindent\textbf{2) Detection False Positive (7/22)}:
\end{revision}
Seven are detection-stage false positives where the function still exists. \textsc{ReCite} correctly recommends no change in five of these.

\begin{revision}
\noindent\textbf{3) Non-Function Entity (6/22)}:
\end{revision}
These cases reference non-function entities (e.g., driver callbacks), which are also noted by \textsc{ReCite}.

% \vspace{0.5em}
% \noindent\textbf{Summary.}
In our 200-case RQ3 evaluation, \textsc{ReCite} generally produces useful repair suggestions.
Off-Target cases (22/200, 11.0\%) are often identifiable through explicit signals such as change-log formatting or \textsc{ReCite}'s own diagnostics.
% Among the 178 actionable cases, nearly half (85/178, 47.8\%) are directly acceptable.
% The remaining partial cases (93/178, 52.2\%), while requiring manual adjustment, contain no misleading content.
\begin{revision}
Among the 178 actionable cases, nearly half (85, 47.8\%) are directly acceptable, while the remaining (93, 52.2\%) are partially acceptable; they still provide helpful guidance for manual repair.  
\end{revision}
% In practice, whether the repair proposes a concrete successor is generally a useful indicator of its readiness for direct application.
We find that whether the repair proposes a concrete successor is generally a useful indicator of its readiness for direct application.
Across the 200 annotated cases, \textsc{ReCite} proposes a concrete successor in 121 repairs. 
Of these, 82 (67.8\%) are directly acceptable (the first two Acceptable subcategories), 34 (28.1\%) are Partial (the first Partial subcategory), and 5 (4.1\%) are Off-Target. 
Regardless of the final label, a concrete successor gives maintainers a clear starting point for evaluating and refining the repair.
% We have submitted 51 \textsc{ReCite}-based patches to the Linux kernel (with manual refinement where needed for Partial cases), of which \important{20} have been accepted and \important{9} more acknowledged by maintainers.
% Combined with 24 manual patches (\important{19} accepted), \textbf{\important{39} of 75 total patches have been accepted and the remainder still under review.}
% We also received strong positive feedback from a senior Linux kernel maintainer: ``\textit{Impressive use of AI and impeccably correctly tagged, this is exactly how I want AI-assisted submissions to look.}''
\begin{revision}
Combined with 24 manual patches, we have submitted 75 repair patches to the Linux community in total.
Their current status is summarized in Table~\ref{tab:patch-outcomes}.
Of these, 50 have been accepted (i.e., queued in maintainer trees or merged into mainline), 3 have received maintainer responses, including 2 with Acked/Reviewed-by but are not yet in any tree, 1 case (\texttt{\_decode\_session6()}, discussed later) is superseded by an alternative refactoring chosen by the developer, and the remainder are still under review.
No patch was rejected for reporting a false issue.
% Among the 51 \textsc{ReCite}-based patches, 31 have been accepted.
To avoid burdening the community, we manually refine \textsc{ReCite}-based patches before submission when needed, typically by simplifying wording or correcting outdated descriptions that \textsc{ReCite} missed.
Among the 31 accepted \textsc{ReCite}-based patches, 15 (48.4\%) are verbatim \textsc{ReCite} patches that required no manual revision.
Together with the 42.5\% Acceptable rate in our manual evaluation, we frame \textsc{ReCite} as a maintainer decision-support tool that produces useful repair candidates, rather than a fully automatic repair system at present.
We also received strong positive feedback from a senior Linux kernel maintainer: ``\textit{Impressive use of AI and impeccably correctly tagged, this is exactly how I want AI-assisted submissions to~look.}''
\end
{revision}

\begin{table}[t]
\centering
\scriptsize
\setlength{\tabcolsep}{4.2pt}
\caption{
\begin{revision}
Outcomes of the 75 submitted Linux kernel patches.
\end{revision}
}
\label{tab:patch-outcomes}
\vspace{-3pt}
\resizebox{\columnwidth}{!}{
\begin{tabular}{cccccc}
\toprule
\textbf{Patch Source} & \textbf{Accepted} & \textbf{Responded} & \textbf{Pending} & \textbf{Rejected} & \textbf{Total} \\
\midrule
Verbatim \textsc{ReCite} patches & 15 & 1 & 7 & 0 & 23 \\
Human-refined \textsc{ReCite} patches & 16 & 2 & 10 & 0 & 28 \\
Manual patches & 19 & 0 & 5 & 0 & 24 \\
\midrule
\textbf{Total} & \textbf{50} & \textbf{3} & \textbf{22} & \textbf{0} & \textbf{75} \\
\bottomrule
\end{tabular}
}
\end{table}

\begin{table}[t]
\centering
\scriptsize
\setlength{\tabcolsep}{2.8pt}
\caption{
(RQ4) Performance on the \textbf{Repair-Worthy Detection 400} dataset.
}
\label{tab:rq4-end-to-end}
\vspace{-3pt}
\begin{tabular}{ccccccccc}
\toprule
\multirow{2}{*}[-0.4em]{\textbf{Sample}}
& \multicolumn{4}{c}{\textbf{\textsc{ReCite}}}
& \multicolumn{4}{c}{\textbf{C4RLLaMA}} \\
\cmidrule(lr){2-5} \cmidrule(lr){6-9}
& \textbf{$P$} & \textbf{$R$} & \textbf{$F1$} & \textbf{$Acc$}
& \textbf{$P$} & \textbf{$R$} & \textbf{$F1$} & \textbf{$Acc$} \\
\midrule
\textsc{ReCite}-Positive ($n{=}200$)
& \textbf{89.0\%} & \textbf{100.0\%} & \textbf{0.942} & \textbf{89.0\%}
& \textbf{90.4\%} & 84.3\% & 0.872 & 78.0\% \\
\textsc{ReCite}-Negative ($n{=}200$)
& -- & 0.0\% & -- & \textbf{99.0\%}
& 1.2\% & \textbf{100.0\%} & 0.023 & 16.5\% \\
Balanced Full ($n{=}400$)
& \textbf{89.0\%} & \textbf{98.9\%} & \textbf{0.937} & \textbf{94.0\%}
& 45.4\% & 84.4\% & 0.590 & 47.2\% \\
% Weighted Overall & \textbf{89.0\%} & 48.4\% & \textbf{0.627} & \textbf{98.9\%} & 2.1\% & \textbf{92.4\%} & 0.041 & 17.1\% \\
\bottomrule
\end{tabular}
\end{table}

% \begin{revision}
% \vspace*{0.1cm}
% \noindent
\textbf{Ablation Results.}
Finally, we compare \textsc{ReCite}  with the rule-based repair baseline and its no-history variant.
The rule-based baseline reproduces \textsc{ReCite}'s repair in only 260 of 869 cases (29.9\%), and even on the simplest pure-name-substitution cases, it identifies the correct successor in just 44 of 71 cases (62.0\%).
The no-history variant is similarly limited:
it matches our original repairs in only 110 of 869 cases (12.6\%), and 9 of 71 (12.7\%) on the same pure-name-substitution subset.
These results show that, in our setting, both LLM-based reasoning and explicit evolution history substantially improve repair over simple rule-based and no-history variants.
% \end{revision}

\subsection{RQ4: End-to-End Detection.}
\label{sec:rq4}

\begin{revision}
Table~\ref{tab:rq4-end-to-end} reports the end-to-end results.
Although \textsc{ReCite} misses two genuinely repair-worthy stale references, it still achieves good end-to-end performance on the balanced full sample ($F1{=}0.937$, $Acc{=}94.0\%$).
Both missed cases were already unresolvable at introduction and thus never entered the tracing and repair steps, i.e., the \emph{True Missing Function} category analyzed in Section~\ref{sec:taxonomy}.
By contrast, C4RLLaMA marks 335/400 (83.8\%) cases as inconsistent and performs poorly on the balanced sample ($F1{=}0.590$, $Acc{=}47.2\%$), suggesting that general CCI methods does not transfer reliably to our setting and may over-flag Linux kernel comments.
\end{revision}

\section{Discussion}
\label{sec:discussion}

\subsection{Categorization of Unresolvable Mentions}
\label{sec:taxonomy}
% For the \verified{7,177 (86.9\%)} mentions that were already unresolvable when the symbol was first written into the comment, we conducted a \emph{manual categorization} to understand their underlying reasons.
% To develop an initial taxonomy, the first author performed open coding on 30 randomly sampled cases, yielding five categories listed below.
% We then sampled an additional 200 cases to form the \textbf{Exist Reason 200} dataset.
% Three other annotators---also undergraduate students majoring in software engineering---independently labeled a pilot batch of 20 cases according to the taxonomy above (with an additional \emph{Other} option for cases not covered), achieving a Fleiss' $\kappa$ of 0.72.
% After resolving disagreements through discussion with the first author, we refined the labeling guidelines accordingly.
% The remaining 180 cases were then divided equally among the three annotators (60 each).
% All 200 annotated cases fall within the five categories established during open coding. No additional categories emerged during the annotation process.
For the \verified{7,177 (86.9\%)} mentions that were already unresolvable when the symbol was first written into the comment, we conducted a manual categorization to understand their underlying reasons.
The first author first performed open coding on 30 randomly sampled cases, establishing the five categories listed below.
\begin{revision}
We then sampled an additional 400 cases to form the \textbf{Exist Reason 400} dataset.
\end{revision}
Three other annotators (also undergraduate students majoring in software engineering) began with a 20-case pilot, achieving a Fleiss' $\kappa$ of 0.72.
After resolving disagreements with the first author, they completed the remaining 380 annotations. 
All 400 annotated cases fell within the five categories, with no additional category emerging.

\begin{enumerate}[itemsep=1pt, topsep=3pt, leftmargin=15pt, label=(\arabic*)]

\begin{revision}
\item \textbf{Non-kernel Entity} (43/400, 10.8\%):
\end{revision}
the symbol refers to an entity not defined in the kernel source, such as a compiler built-in, a C library function, or a function from an external project.
For example, the comment \textit{``CONFIG\_UBSAN\_TRAP inserts a UD2 when it sees \textbf{\_\_builtin\_unreachable()}''} refers to a GCC compiler built-in, not a kernel function.

\begin{revision}
\item \textbf{Kernel Concept, Not a Function} (153/400, 38.3\%):
\end{revision}
the symbol refers to a kernel entity but does not refer to a function.
% For example, in the comment \textit{``arg: passed to rmap\_one() and \textbf{invalid\_vma()}''}, the ``\texttt{invalid\_vma}'' is a function-pointer field in \texttt{struct rmap\_walk\_control}, not a standalone function.
% Other instances include ``\textit{kexec()}'' used as a verb meaning \textit{``perform a kexec operation''} and ``\textit{cpa()}'' as an abbreviation for \textit{``Change Page Attributes''}.
For example, in comments, ``\textit{invalid\_vma()}'' may denote a function-pointer field rather than a standalone function, and ``\textit{kexec()}'' can be used as a verb meaning \textit{``perform a kexec operation''}.

\begin{revision}
\item \textbf{Abbreviation/Wildcard} (113/400, 28.3\%):
\end{revision}
the symbol refers to one or more real kernel functions, but the comment uses an abbreviation or suffix pattern that does not exactly match any function name.
For example, the comment \textit{``completion is now handled in the \textbf{regular\_isr()}''} refers to the function \texttt{wmt\_mci\_regular\_isr()} in the same file, but resolving the abbreviated name \textit{regular\_isr} alone yields no match.
% Another pattern is wildcard-style references such as \textit{``octeon\_irq\_ip\{2,3\}\_ciu()''}, which refers to multiple functions simultaneously.

\begin{revision}
\item \textbf{True Missing Function} (69/400, 17.3\%):
\end{revision}
the symbol refers to a specific kernel function that genuinely does not exist at the queried version.
We further sub-categorize these cases:

(a) \textbf{Typo} (33/69): the symbol is a misspelling of an existing function, e.g., ``\textit{cmpxcgh()}'' for the actual function \texttt{cmpxchg()}.
(b) \textbf{Already Obsolete at Introduction} (32/68): the referenced function had already been removed or renamed before the comment was written.
For example, the function \texttt{plug\_ctx\_cmp()} was refactored into \texttt{plug\_rq\_cmp()} in 2018, but a comment created in 2021 still referenced the obsolete name---likely due to copy-paste from an earlier source.

(c) \textbf{Explicitly Declared Absence} (4/69): the comment itself makes clear that the referenced function does not exist, e.g., in a TODO note.

% \begin{enumerate}[itemsep=1pt, topsep=3pt, leftmargin=15pt, label=(\alph*)]

% \item 
% \begin{revision}
% \textbf{Typo} (33/69): 
% \end{revision}
% the symbol is a misspelling of an existing function, e.g., ``\textit{cmpxcgh()}'' for the actual function \texttt{cmpxchg()}.

% \item 
% \begin{revision}
% \textbf{Already Obsolete at Introduction} (32/68): 
% \end{revision}
% the referenced function had already been removed or renamed before the comment was written.
% For example, the function \texttt{plug\_ctx\\\_cmp()} was refactored into \texttt{plug\_rq\_cmp()} in 2018, but a comment created in 2021 still referenced the obsolete name---likely due to copy-paste from an earlier source.

% \item 
% \begin{revision}
% \textbf{Explicitly Declared Absence} (4/69):
% \end{revision}
% the comment itself makes clear that the referenced function does not exist, e.g., in a TODO note.
%  % e.g., \textit{`` FIXME: \ldots this also needs to be worked with Christoph, \textbf{register\_NMI\_handler()}''}.
% \end{enumerate}

\begin{revision}
\item \textbf{Tool False Positive} (22/400, 5.5\%):
\end{revision}
the function exists at the queried version but was missed by our tool.

\end{enumerate}

% The categorization reveals that identifying repair-worthy stale function references goes well beyond detecting unresolved symbols.
% Further discussions about identification and repair are as follows.

\subsection{Challenges and Insights}

As discussed above, both identifying and repairing stale function references are considerably more complex than initially anticipated.
We summarize the key challenges and insights below.

% \vspace{0.5em}
% \noindent
\textbf{Challenges in identification.}
The difficulty of identifying repair-worthy stale references stems from two sources: the diversity of natural language in comments, and the complexity of the code itself.
\textbf{On the language side}, developers frequently use function-form symbols to refer to entities that are not kernel-managed functions---compiler built-ins, C library routines (Cat.~1), struct fields, or verbs (Cat.~2).
These do not require function-level repair.
Beyond non-targets, natural language may also complicate what a symbol actually refers to.
Developers may use abbreviations or wildcard-style names (Cat.~3), which turn a one-to-one symbol-to-entity mapping into a one-to-many relationship.
And the abbreviation could also become stale.
For instance, \texttt{pci\_dma\_sync()} is a shorthand for a family of PCI DMA synchronization functions (e.g. \texttt{pci\_dma\_sync\_\{single\_for\_cpu, sg\_for\_device\}()}, etc.), all of which were removed in commit \texttt{7968778} and replaced by the unified \texttt{dma\_sync\_*} API.
Moreover, typos and omitted parentheses (e.g., writing \texttt{foo} instead of \texttt{foo()}) could further increase ambiguity and cause references to escape our extraction.
\textbf{On the code side}, even when a matching function definition exists, the specific reference may still be stale.
For instance, if a comment in an ARM source file mentions \texttt{platform\_cpu\_die()} but the only matching definition is found under a different architecture (i.e., \texttt{arch/blackfin/}), the reference is still outdated since it clearly does not intend to point there.
Such cases (\texttt{NO\_MATCHED\_DEFINITION} in our entity selection) are effectively a form of \texttt{Already Obsolete at Introduction}, requiring architecture-aware reasoning to disambiguate.
Given these complexities, we do not claim to have mapped all stale function references in the Linux kernel.
Our approach---anchoring on ``\texttt{symbol()}''-form tokens---provides a practical starting point, and we focus repair efforts on the subset where a source entity can be located at the symbol-introducing commit.

% \vspace{0.5em}
% \noindent
\textbf{Challenges in repair.}
Tracing a stale reference back to its source entity is a prerequisite for history-grounded repair, yet this cannot always be achieved.
The Linux kernel has been under development since 1991, but it only adopted Git in 2005, leaving pre-Git history difficult to access~\cite{lyu2024evaluating}.
Moreover, comment authors do not always reference valid entities at the time of writing (e.g., through copy-paste from older code).
Combined with non-kernel and non-function references (Cat.~1--3), 7,177 of 8,258 mentions (86.9\%) are already unresolvable at introduction.
In this work, we focus on the 1,081 resolvable mentions, where a source entity can be located---this also naturally filters out non-kernel entities and other noise, letting us concentrate on cases with clear repair value.
% In this work, we focus on the 1,081 resolvable mentions, where a source entity can be located---this naturally filters out non-kernel entities and non-function references, though a small fraction of the filtered mentions (e.g., Cat.~4a, already obsolete at introduction) are genuine stale references that our method currently cannot trace.
Even for these traceable cases, repairing a stale reference is far from simple name substitution.
A function's removal or refactoring often reflects deeper changes in code logic.
For example, in the comment \textit{``It must be done after \texttt{xfrm6\_policy\_check()}, because \textbf{\_decode\_session6()} uses \texttt{IP6CB()}''}, \texttt{\_decode\_session6()} was first renamed to \texttt{decode\_session6()} and later refactored to no longer use \texttt{IP6CB()} (commit \texttt{7a02070}).
As a result, the ordering constraint itself no longer holds, and the entire sentence---not just the function name---should be removed.
We submitted this repair as a kernel patch, but the maintainer's response went even further, questioning whether the removal of this constraint makes the repeated \texttt{fill\_cb}/\texttt{restore\_cb} call pattern it necessitated redundant as well.
This suggests that stale comments can signal deeper code-level technical debt.
Such cases call for repair strategies that go beyond lexical replacement.
In future work, we plan to explore more flexible, agent-based approaches~\cite{lyu2026practitioners} to address them, including handling typos and wildcard-style references.

\section{Threats to Validity}
\label{sec:threats}

\noindent
\textbf{Internal validity.}
Our $\textsc{Resolve}$-Coccinelle pipeline operates on unexpanded source text, so it cannot recognize function definitions generated through C preprocessor macros.
Coccinelle can resolve this by performing full preprocessing with the kernel build configuration, but this drastically increases analysis time and limits results to the single hardware architecture selected in the config file.
Operating on unexpanded source text allows our pipeline to find definitions for all architectures.
This trade-off favors precision over recall ($P{=}96.7\%$): $\textsc{Resolve}$-Coccinelle may over-report unresolved symbols, but rarely misses a genuinely resolved one.
For our task, this is the safer direction---over-reported cases are easily filtered during evolution tracing, whereas a missed stale reference would silently forgo repair.
% Another internal threat arises from the use of an LLM (DeepSeek-V3.2) in entity selection, refactoring analysis, and repair generation.
% Results may vary with different models or parameter settings. We mitigate this by setting the temperature to 0 and applying validation-and-retry at each stage to guard against hallucination.
%%NOTE:Since we will be adding more data, the experimental results below might change.
\begin{revision}
Another threat is that we track the function changes commit-by-commit, so a refactoring spread across commits (e.g., delete then re-add) may be misclassified as \emph{Removed}.
To assess this, we sampled 50 cases flagged as \emph{removed} and compared each deleted body against later-added functions, treating $\geq$ 50\% identifier-token overlap (Jaccard $\geq$ 0.5) as a possible multi-step refactoring; none was detected.
Further threats stem from our hyperparameters and choice of LLM.
For identifying the symbol-introducing commit, we use Jaccard $\geq 0.7$ to match the symbol-bearing comment; rerunning this step at 0.6 and 0.8 changed only 0.86\% / 0.85\% of commits, suggesting low sensitivity to this threshold.
% CITE: lyu2024evaluating
For the LLM, we use DeepSeek-V3.2 as a cost-effective choice with temperature 0 for output stability, following prior SE studies~\cite{chen2025linebreaker, ouyang2025empirical}.
A stronger model shows better performance, e.g. on the RQ2 Entity Selection 200 dataset, Claude Opus 4.7~\cite{Claude} improves entity-detection $F1$ from 0.964 to 0.975 and file-path match accuracy from 98.3\% to 99.4\%.
This suggests that our results are not tied to the strongest available model and may improve with stronger LLMs.
Our prompts use only structured inputs (no Linux-specific few-shot examples), though their effectiveness on other projects requires further evaluation.
\end{revision}

\vspace{0.5em}
\noindent
\textbf{External validity.}
We evaluate \textsc{ReCite} solely on the Linux kernel.
Given its scale and complexity, we expect the approach to generalize to other C projects with a Git history.
While \textsc{ReCite}'s core idea of static-analysis-assisted LLM reasoning is language-agnostic, its effectiveness on other languages remains to be validated.

\vspace{0.5em}
\noindent
\textbf{Construct validity.}
All our annotated datasets meet or exceed the minimum sample size for a 95\% confidence level with a $\pm$10\% margin of error.
% ### Disproportionate Stratified Sampling Version：
% For Symbol Resolution 150 and Refactoring Analysis 150 dataset, we used disproportionate stratified sampling, as disagreement cases are rarer but more informative for tool comparison.
% ### Original Version：
\begin{revision}
Two of them---\textbf{Symbol Resolution 200} and \textbf{Refactoring Analysis 200}---are sampled from cases where tools disagree, which may introduce sampling bias.
To verify that tool performance generalizes, we additionally sampled 100 agreement cases for each dataset.
On Symbol Resolution, all five $\textsc{Resolve}$ variants achieve 100\% accuracy.
On Refactoring Analysis, \textsc{ReCite} achieves \verified{$F1{=}0.891$} on binary classification and $F1{=}1.00$ on successor identification.
The lower binary $F1$ compared to the disagreement set (\verified{0.974}) is attributable to 9 borderline cases where the function was deleted but its functionality was merged by another one, leading both tools to label them as \emph{removed} while annotators favor \emph{refactored}.
The perfect successor identification (\verified{$F1{=}1.00$}) further demonstrates that \textsc{ReCite} reliably traces function evolution on both agreement and disagreement cases.
\end{revision}
Similarly, given the rarity of stale references, RQ4 uses balanced rather than purely random sampling, which may limit our view of their true prevalence across all 83,241 mentions. Nevertheless, on the annotated sample, \textsc{ReCite} achieves high precision with only two false negatives.
% Additionally, to ensure annotation reliability, all datasets follow a pilot-then-split protocol with the following three steps: (1) two annotators independently label a shared pilot batch (20 cases); (2) disagreements are resolved through discussion with the first author; and (3) the remaining cases are divided between annotators.
% Inter-annotator agreement across pilot batches reaches at least substantial agreement (minimum $\kappa$ = \verified{0.62}), supporting the reliability of our ground truth.
Additionally, to ensure annotation reliability of our ground truth, all datasets follow a pilot-then-split protocol, 
and the inter-annotator agreement across pilot batches reaches at least substantial agreement (minimum $\kappa$ = \verified{0.62}).
\begin{revision}
Finally, Section~\ref{sec:taxonomy} shows that an unresolved symbol (i.e., one without a matching function) is not necessarily a stale reference requiring repair.
And we focus on staleness in the narrow sense of function references, specifically cases where the referenced function has disappeared through code evolution. Cases involving other code entities (e.g. variables), or semantic drift while the referenced function will be our future work.
Lastly, given the 42.5\% Acceptable rate in RQ3, we position \textsc{ReCite} as a decision-support tool rather than a fully automatic repair system, at present.
\end{revision}

\section{Conclusion and Future Work}
We presented \textsc{ReCite}, a three-stage approach for automatically identifying and repairing stale function references in Linux kernel comments.
On kernel v6.18-rc1, \textsc{ReCite} extracted 83,241 function-form mentions and detected 869 stale references with generated repair suggestions.
A manual evaluation on 200 sampled repairs shows that 85 (42.5\%) generated suggestions are directly applicable.
To date, \important{50} of 75 our submitted patches have been accepted.

In future work, we plan to broaden coverage by handling abbreviations, wildcard-style references, and omitted parentheses.
We also aim to explore agent-based approaches that can reason about deeper semantic changes, and to investigate the generalizability of our approach to other large, long-lived C codebases.

\begin{acks}
Yucong Guan, Jiaqi Sun, Yifeng Pan, Zhenyan Song, and Yuxuan Li also contribute to this work.
This work was supported in part by the Fundamental and Interdisciplinary Disciplines Breakthrough Plan of the Ministry of Education of China under Grant JYB2025XDXM118; 
in part by the National Natural Science Foundation of China under Grant 62572237, Grant 62302210, and Grant 72371125; 
in part by the 2024 Development and Testing Tools Project under Grant CEIEC-2024-ZM02-0066; 
in part by the Frontier Technologies R\&D Program of Jiangsu under Grant BF2024059; 
in part by the Natural Science Foundation of Jiangsu Province under Grant BK20241195;
and in part by the PTCC project SWHSec under Grant ANR-22-PTCC-0001.
This research / project is supported by the National Research Foundation, under its Investigatorship Grant (NRF-NRFI08-2022-0002). 
Any opinions, findings and conclusions or recommendations expressed in this material are those of the author(s) and do not reflect the views of National Research Foundation, Singapore.
% Hongyu Kuang and Yunbo Lyu are the corresponding authors.
\end{acks}

\section*{Data Availability Statement}
Our replication package, including all datasets, scripts, and prompts, is available at \important{\url{https://doi.org/10.5281/zenodo.19232503}}.


\balance{}

\bibliographystyle{ACM-Reference-Format}
\bibliography{reference}

\end{document}